\documentclass{article}

\usepackage{PRIMEarxiv}

\usepackage[utf8]{inputenc} 
\usepackage[T1]{fontenc}    
\usepackage{hyperref}       
\usepackage{url}            
\usepackage{booktabs}       
\usepackage{amsfonts}       
\usepackage{nicefrac}       
\usepackage{microtype}      
\usepackage{lipsum}
\usepackage{fancyhdr}       
\usepackage{graphicx}       
\graphicspath{{media/}}     
\usepackage{xspace} 
\usepackage{hyphenat}
\usepackage{siunitx}
 \usepackage{amsmath}
 \usepackage{multirow}
\usepackage{subfigure}
\usepackage{appendix}
\title{Translating the internal climate variability from climate variables to hydropower production
}

\author{
  Divya Upadhyay \\
  Discipline of Civil Engineering, \\
  Indian Institute of Technology Gandhinagar, India \\
  \And
  Sudhanshu Dixit \\
  Discipline of Civil Engineering, \\
  Indian Institute of Technology Gandhinagar, India \\
  \And
  Udit Bhatia \\
  Discipline of Civil Engineering, \\
  Indian Institute of Technology Gandhinagar, India \\
  \texttt{bhatia.u@iitgn.ac.in}
}

\begin{document}
\maketitle

\begin{abstract}
Quantifying uncertainties in estimating future hydropower production directly or indirectly affects India's energy security, planning, and management. The chaotic and nonlinear nature of atmospheric processes results in considerable Internal Climate Variability (ICV) for future projections of climate variables. Multiple initial condition ensembles and multi-model ensembles of earth system models are often used to analyze the role of ICV and model uncertainty in precipitation and temperature. However, there are limited studies focusing on quantifying the role of internal variability on impact variables, including hydropower production. In this study, we analyze the role of internal variability and model uncertainty on three prominent hydropower plants of India using multiple initial condition ensembles of the EC-Earth3 and multiple models from CMIP6. We estimate the streamflow projections for all ensembles using the Variable Infiltration Capacity (VIC) hydrological model for four time periods, historical (1985-2014), near-term (2015-2044), mid-term (2045-2074) and far-term (2075-2100). We estimate maximum hydropower production generated using monthly release and hydraulic head available at the reservoir. We also analyzed the role of bias correction in hydropower production. The results show that ICV plays a significant role in estimating streamflow and hydropower estimation for monsoon and throughout the year, respectively. Model uncertainty contributes more to total uncertainty than ICV in estimating the streamflow and potential hydropower. However, ICV is increasing towards the far-term. We also show that bias correction does not preserve the internal variability in estimating the streamflow. Although there is an increase in uncertainty for estimated streamflow, mean hydropower shows the decrease towards the far-term for February to May, more prominent for MICE than MME. The results suggest a need to incorporate uncertainty due to internal variability for addressing power security in changing climate scenarios.
\end{abstract}


\section{Introduction}
Sustaining an optimal reservoir operation under various uncertainties is challenging in changing climate for hydropower production. The increasing population, urbanization and industrialization raise the demand for the energy consumption of fossil fuels, water, material, and available natural resources, which further poses the issue of energy security and clean environment \cite{madlener2011impacts, saraswat2021empirical}. The large energy requirement in India is fulfilled by non-renewable sources, which is fast depleting \cite{sharma2021opportunities}. Hydropower is a vital source of renewable energy, which provides over 72 \% of all renewable energy globally \cite{gernaat2017high}. However, only about 17 \% of hydropower potential of 150,000 MW has been tapped in India \cite{kumar2010renewable}. For India, hydropower and small hydropower are the second-highest contributors towards renewable energy consumption, which is close to 12 \% of the total capacity \cite{sharma2013comprehensive}. The impacts of climate change on hydropower production affect energy generation in the near term and by the end of the 21st century \cite{lehner2005impact, liu2016projected, turner2017examining, ali2018projected}. Thus, tremendous opportunities in hydropower generation and its future expansion exist, which help us meet the soaring energy demands. However, hydropower production is directly related to the water availability in reservoirs that depends on streamflow, which is characterized by uncertainties of the processes that result in streamflow \cite{lehner2005impact}. The uncertainty associated with precipitation, temperature and wind pattern of a region affects the streamflow uncertainty \cite{redmond1991surface, shook2015transformation}. \cite{biemans2009effects} have analyzed the effects of precipitation uncertainty on discharge calculations for 294 river basins worldwide and shown that average precipitation uncertainty (about 30 \%) can lead to higher uncertainty in discharge (about 90 \%). Quantification of uncertainties associated with future projections of streamflow is a critical challenge for climate change impact studies \cite{xu2004review} as it necessitates obtaining the range of possible future variations in streamflow projections \cite{hingray2019climate}. 

Characterizing and quantifying uncertainty projection of streamflow and hydropower potential considering the climate change projections is important to plan for adaptation and mitigation \cite{deser2012uncertainty}. It is essential to quantify and discuss the role of different uncertainties and their propagation to future hydrological projections as it is a fundamental characteristic of prediction \cite{finger2012projections}. The estimated variables' uncertainty sources are forcing uncertainty, model uncertainty and internal climate variability \cite{hawkins2011potential, deser2012uncertainty, topal2020refining}. The relative importance of different uncertainty sources varies depending on the type of variable, temporal and spatial scales, and specifically its nature \cite{hawkins2011potential, deser2012uncertainty, gao2020assessment}. For example, model uncertainty reflects our lack of knowledge or inability to encapsulate the existing knowledge within the climate models, and it is considered as potentially reducible as models improve \cite{deser2012uncertainty}. The Multi-Model Ensembles (MMEs) are generally used to estimate the model uncertainty \cite{singh2020sensitivity}. Whereas, Internal Climate Variability (ICV), which refers to natural variability that arises from processes in the coupled land, biosphere, ocean, atmosphere, and cryosphere system in the absence of external forcing \cite{deser2020insights, hyun2020understanding}, considered as irreducible \cite{madden1976estimates, hawkins2009potential, deser2012communication, fischer2014models, bhatia2019precipitation, lehner2020partitioning}. ICV dominates climate uncertainty over decadal prediction horizons at regional to local scales, where stakeholders are more interested \cite{kumar2018intercomparison}. It is typically handled by considering Multiple Initial Condition Ensembles (MICE), assuming that it satisfactorily captures the internal variability \cite{deser2020insights, upadhyay2021depth}. The MICE are generated by applying minor perturbations to the initial state of the model such that the different climate projections behave as surrogates of climate variability \cite{deser2012communication, stocker2013climate, asch2016demystifying, kumar2018intercomparison, innocenti2019projected}.  

The importance of ICV has received considerable attention recently from scientific community  \cite{deser2012uncertainty, deser2014projecting, deser2020insights, lehner2020partitioning, bhatia2019precipitation, upadhyay2021depth}. Several studies have explored the role of internal variability to assess future climate outcomes such as surface air temperature, a vertical profile of recent tropical temperature trends and precipitation \cite{deser2012uncertainty, deser2012communication, deser2014projecting, deser2020insights, lehner2020partitioning, mitchell2020vertical}. A few studies have quantified role of ICV in estimating  streamflow \cite{fatichi2014does, champagne2020future}, extreme precipitation \cite{bhatia2019precipitation, upadhyay2021depth}, sea-level rise \cite{tsai2020role}, air quality and associated health risks in a warming world \cite{saari2019effect}. \cite{mankin2020value} have criticised initial condition large ensembles as they are resource-intensive, redundant, and biased. Despite this criticism, they have shown that large ensembles provide unique information consistent with the insights and support robust adaptation decision‐making regarding freshwater resources. Climate variability may cause large uncertainties in climate at regional scale  \cite{hawkins2011potential, deser2012uncertainty, deser2014projecting}. Recently, \cite{upadhyay2021depth} have quantified the relative contribution of uncertainty due to ICV and model uncertainty in the depth and volatility of Indian summer monsoon rainfall extremes. They have shown that ICV is comparable and even higher than model uncertainty in estimating extreme precipitation indices in central India. However, the understanding role of ICV in various meteorological variables and its translation to streamflow and hydropower estimation has received relatively less attention for India. A few researchers have considered the translation of different uncertainties of climate variables to hydropower and analyzed it using hydrological models \cite{poulin2011uncertainty, finger2012projections, oyerinde2016quantifying}. Here, we consider the uncertainties associated with precipitation, maximum temperature ($T_{max}$), minimum temperature ($T_{min}$), wind and its impact on streamflow estimation and subsequently on hydropower production, as the range of uncertainty in input data, has a significant influence on the output, which may not be neglected in the communication of results \cite{biemans2009effects}.

Several studies have analyzed the impact of climate change on hydropower production and shown that it affects energy generation in the near term and by the end of the 21st century \cite{lehner2005impact, liu2016projected, turner2017examining, van2016power, ali2018projected}. \cite{carvajal2017assessing} have assessed the sensitivity of hydropower production using 40 multi-model ensembles and shown that annual hydroelectric power production ranges between $- 55$ and +39 \% of the mean historical outputs for the period 2071–2100. For seven large hydropower projects of India, \cite{ali2018projected} have assessed climate change impacts on hydropower production and analyzed model uncertainties and projected increase in mean hydropower production and annual streamflow in non-snow dominated areas. Similarly, \cite{caceres2021hydropower} have evaluated the impact of climate on hydropower using 21 GCMs and quantified the role of model uncertainty of 134 hydropower plants of South America. However, studies on the role of ICV in estimating hydropower has rarely been studied using large ensembles. For example, \cite{finger2012projections} have examined uncertainties due to internal variability, which is intrinsically sampled by assessing the interannual variability within the projected 28-year duration. The role of the ICV using large ensembles in estimating hydropower production in India is still unexplored, despite exploring the impact of climate change on hydropower production. 

This study attempts to translate the meteorological variables' uncertainties due to ICV and model uncertainty to hydropower production at a regional scale. We analyze the role of model uncertainty and ICV in estimating streamflow and hydropower generation. We analyze the effect of bias correction of meteorological variables such as precipitation, $T_{max}$, $T_{min}$ and wind data on streamflow estimation and its translation to hydropower production estimation. Our study may help scientists and policymakers to understand and communicate the role of internal variability at a regional scale and provide a way to assimilate multiple sources of information to justify actions in climate change adaptation.

The following section of the paper shows the study area and data set used for the analysis, followed by the methodology section. In the methodology section, we first discuss a flowchart for uncertainty analysis. Then, the subsection provides details of streamflow and potential hydropower estimation. We discuss the uncertainty analysis in this paper's result and discussion section.

\section{Study area and data}

\begin{figure}
\begin{center}
    \fbox{\includegraphics[width=0.9\textwidth,keepaspectratio]{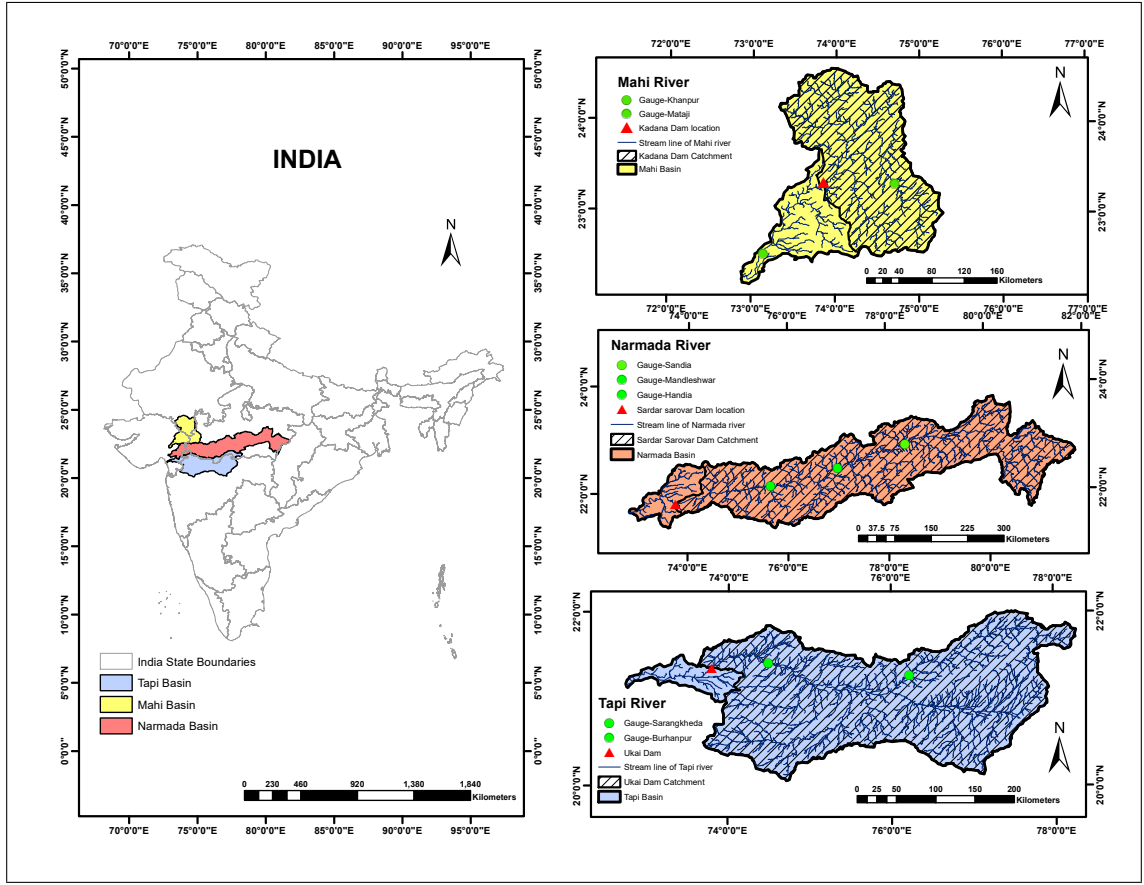}}
    \caption{\textbf{Study Area}: Location of three basins of India are shown with the basin watershed. The red triangle shows the dam's location, and the green circle shows the location of gauges considered for calibration and validation of the VIC hydrological model. Mahi, Narmada and Tapi basins are given from top to bottom in the second column, respectively. The shaded part represents the catchment area contributing to the dam location.}
    \label{Fig1}
    \end{center}
\end{figure}

India has experienced significant changes in temperature and precipitation over the past few decades, which is likely to change by the end of the twenty-first century \cite{mishra2016hydrologic, mohammad2019temperature, almazroui2020projections}. India is the fifth-largest producer of hydroelectric power globally, with a 50.07 GW capacity \cite{hydro2021}. India's power grid is divided into northern, western, eastern, southern, and northeastern regions \cite{halder2013comprehensive}. The western region covers Maharashtra, Gujarat, Madhya Pradesh, and Goa states of India, with the highest capacity for hydroelectric potential \cite{jain2007hydrology, halder2013comprehensive, sharma2020impetus}. This study considers the three major hydropower plants such as Sardar Sarovar, Ukai, and Kadana, located in Gujarat, India. The details of these hydropower plants are given in the appendix (Table \ref{Details of powerplant}), and more details can be obtained from India Water Resources Information System (India-WRIS) (\url{https://indiawris.gov.in/wris/}). Gujarat is the fifth largest Indian state located in the western region of India and the ninth largest state by population with approximately 60.4 million \cite{census2011}. The study area and location of the three basins are shown in Figure \ref{Fig1}. We consider the hydropower plants with greater than 100 MW capacity for our analysis. The Sardar Sarovar Project has six River Bed Power House (RBPH) with 200 MW capacity and five Canal Head Power House (CHPH) with 50 MW capacity \cite{sahoo2014sardar}. The Ukai and Kadana dam has hydropower with an installed capacity of 305 MW and 240 MW, respectively \cite{ghose2019development, sharma2018mahi}.

We use 50 Multiple Initial Condition Ensembles of the CMIP6-EC-Earth3 to analyze the role of internal variability, which are available from the Earth System Grid Federation \url{(https://esgf-data.dkrz.de/projects/esgf-dkrz/)} \cite{hazeleger2012ec, bilbao2021assessment}. We use eight Multi-Model Ensembles from Climate Model Intercomparison Project Phase-6 (CMIP6) to analyze model uncertainty. The models considered for the model uncertainty analysis are ACCESS-CM2, CanESM5, EC-EARTH3, GFDL-ESM4, INM-CM4-8, INM-CM5-0, MRI-ESM2-0, NorESM2-LM, which are available from World Climate Research Program Coupled Model Intercomparison Project \url{(https://esgf-node.llnl.gov/projects/cmip6/)}. The resolution of each model is given in the appendix (Table \ref{List of CMIP6 data}). We consider low-to-moderate emission scenarios Shared Socioeconomic Pathway (SSP) SSP2-4.5 (approximately corresponding to the RCP-4.5 scenario) for the analysis \cite{meinshausen2020shared, eyring2016overview}. We selected models from CMIP6 based on data availability of precipitation, maximum temperature ($T_{max}$), minimum temperature ($T_{min}$) and wind for the historical period (1985–2014) and the future period (2015–2100) under SSP2-4.5 (SSP245). We use one initial condition realization (r1i1p1) to analyze model uncertainty. 

For hydrological modelling, we use the observed daily gridded precipitation, minimum and maximum temperature from the India Meteorological Department (IMD) \cite{pai2015analysis} at \ang{0.25} X \ang{0.25} resolution. Observed daily wind speed data were obtained from \cite{sheffield2006development, ali2018projected} at \ang{0.25} X \ang{0.25} resolution. All model (MICE/MME) data are interpolated at \ang{0.25} x \ang{0.25} gridded daily precipitation, $T_{max}$, $T_{min}$ and wind dataset over Narmada, Tapi and Mahi basins using linear interpolation. We use observed monthly streamflow data from India-WRIS to calibrate and validate Variable Infiltration Capacity (VIC) Macroscale Hydrologic Model \cite{liang1994simple}. We used vegetation parameters from advanced very high-resolution radiometer global land cover information at a 1-km spatial resolution \cite{hansen2000global, tiwari2019prediction, sheffield2007characteristics} and soil data from the Harmonized World Soil Database, version 1.2 (\url{http://www.fao.org/soils-portal/soil-survey/soil-maps-and-databases/harmonized-world-soil-database-v12/en/}). The storage and height of the reservoir are obtained from Sardar Sarovar Narmada Nigam Ltd. \url{(https://sardarsarovardam.org/hydrology.aspx)} for Sardar Sarovar reservoir. For the ukai and kadana reservoir, we obtained data of storage and height relationship from India-WRIS to compute hydropower potential. 

\begin{figure}
\begin{center}
    \fbox{\includegraphics[width=0.9\textwidth,keepaspectratio]{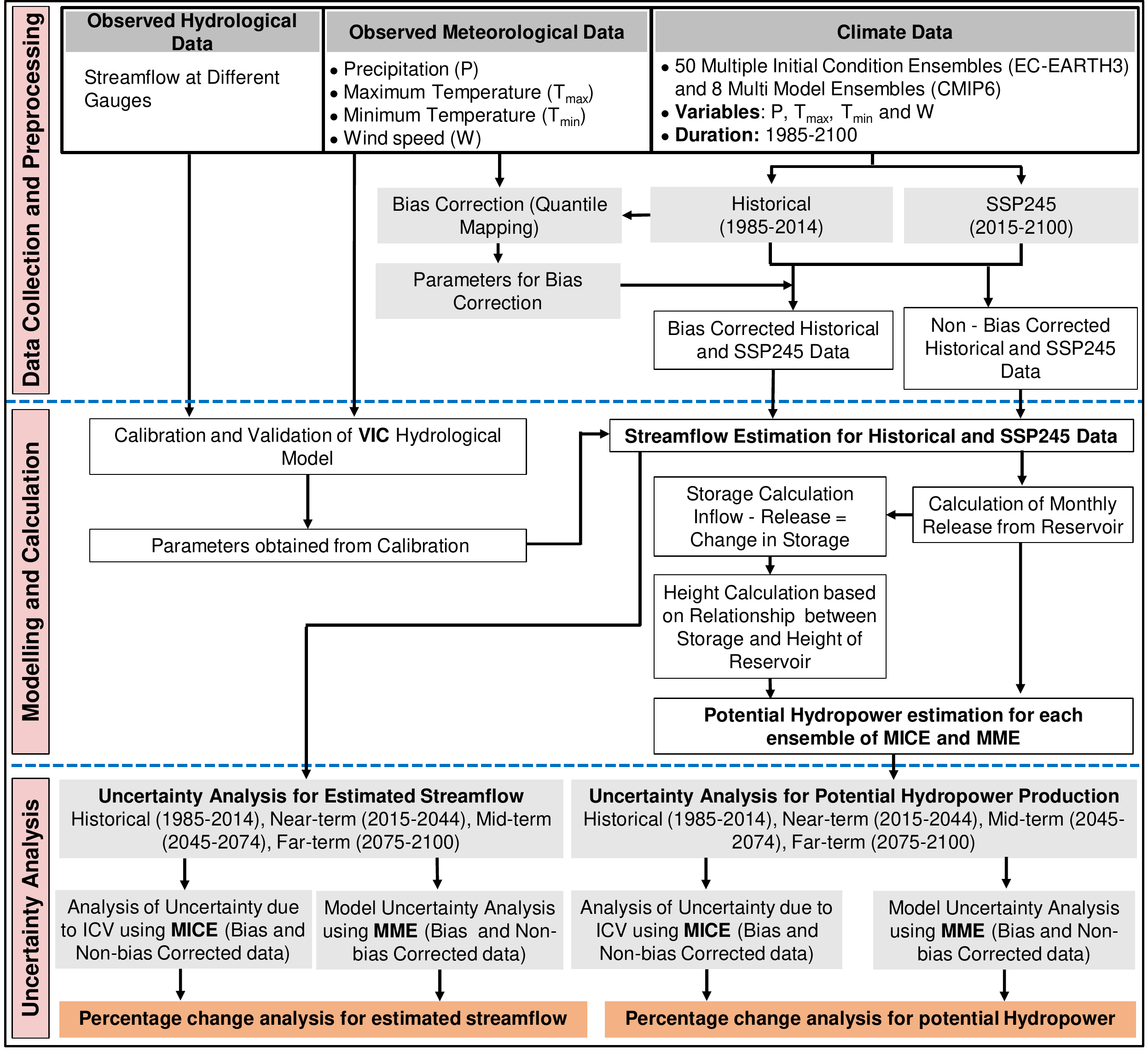}}
    \caption{\textbf{Flowchart of the study}: Quantification of uncertainty arising from internal variability and model uncertainties in the estimation of streamflow and hydropower production, using different steps involved including bias correction, streamflow estimation, hydropower estimation and uncertainty analysis.}
    \label{Fig2}
    \end{center}
\end{figure}

We have analyzed the role of ICV and model uncertainty in estimating streamflow and hydropower production for three hydropower plants. The methodology considered for the proposed work is as shown in Figure \ref{Fig2}, divided into three sections. 

First, the data collection and preprocessing section show the different data used for the analysis. We use observed precipitation (P), $T_{max}$, $T_{min}$, wind (W) and streamflow of different gauges as shown in Figure \ref{Fig1} for calibration and validation of the VIC hydrological model. We use Earth System Models (ESMs) outputs (P, $T_{max}$, $T_{min}$ and W) for estimation of streamflow and potential hydropower. ESM outputs have large systematic biases relative to observational data sets owing to the imperfect representation of the model's physical processes {\cite{grillakis2017method, upadhyay2021depth}}. Thus, it should not be directly used to develop climate change adaptation policies without some form of prior bias correction {\cite{sharma2007spatial,  piani2010statistical, grillakis2017method}}. The objective of bias correction is to adjust the statistical properties of climate simulations with observations. Several studies have used Quantile Mapping (QM) bias correction and shown that it substantially reduces the biases  {\cite{maraun2013bias, cannon2015bias, ngai2017bias, harilal2021augmented, tiwari2021influence}}. Recently, {\cite{chen2021climate}} have compared bias-corrected climate model outputs and post-process model output for streamflow estimation and recommended use of bias correction. {\cite{haddeland2012effects}} have analyzed the effects of bias correction of radiation, humidity and wind estimates on evapotranspiration and runoff estimates and shown that estimates are relatively similar. Differences between simulated and observed radiation, humidity and wind values are smaller than for temperature and precipitation for the bias and non-bias corrected hydrological projections. Bias correction may affect the consistency between the ESM output variables such as P, $T_{max}$, $T_{min}$ and W  {\cite{wang2009multimodel, haddeland2012effects, ehret2012hess, muerth2013need}}. The use of bias-corrected data is an open question and unresolved issue among scientists, especially for climate change impact studies such as hydropower production and agricultural production  {\cite{ehret2012hess, hempel2013trend, laux2021bias}}. Thus, the effect of bias correction on estimated streamflow and potential hydropower is explored in this study.  

We use observed and climate data for each variable such as daily P, $T_{max}$, $T_{min}$, and W of each ensemble (MICE/MME) to analyze the effect of bias correction. We perform quantile-quantile mapping  {\cite{maraun2013bias}} for bias correction for each variable separately. We use the 'qmap' library of Rpy2 in python for data from 1985 to 2014 and try to match quantiles at intervals of 0.01. We estimate parameters for bias correction using observed and historical ensemble data (1985-2014). These parameters are then applied to historical and SSP245 data to obtain bias-corrected historical and SSP245 data. We perform analyses for bias and non-bias corrected data separately. 

The modelling and calculation section shows the flowchart to estimate streamflow and potential hydropower production. First, we calibrate and validate the VIC hydrological model using different gauge streamflow data to obtain calibrated hydrological parameters. Streamflow estimated using calibrated parameters and climate data (bias and non-bias corrected data). These streamflow estimates and storage-height relationships are used to calculate potential hydropower. The last section of the flowchart shows the data and duration used for uncertainty analysis. The details about streamflow estimation and potential hydropower estimation are given below. 

\subsection{Streamflow estimation}
We estimate streamflow using the VIC hydrological model {\cite{liang1994simple}} at the reservoir for three hydropower plants. The VIC is a semi-distributed model developed based on energy and water balance equations solved for individual grid cells. The VIC generate baseflow and streamflow using Arno model conceptualization  {\cite{todini1996arno}} and the infiltration mechanism utilized in the  {\cite{ren1992xinanjiang}}. We use observed daily data of precipitation, $T_{max}$, $T_{min}$, wind data vegetation parameters, soil parameter files, and observed streamflow data to calibrate and validate the model. In the VIC model, the assumption is that water can only enter through the atmosphere, and water entering the channel can not flow back to the soil  {\cite{dang2020software}}. The observed meteorological forcing data such as P, $T_{max}$, $T_{min}$, and W are fed as input data. We first simulate baseflow and surface runoff from each grid cell. Then, runoff from individual cells is routed to gauge location using the routing model, developed by  {\cite{lohmann1996large}}. Channel routing based on the linearized Saint-Venant equation is used to simulate the discharge at the gauge location by assuming that all runoff exits a cell flows in a single flow direction  {\cite{gao2010water, zhang2017flash}}. 

We have calibrated and validated the model using the Handia, Mandleshwar and Sandia gauge station for the Narmada basin. Similarly, we consider Sarangkheda and Burhanpur for the Tapi basin; Mataji and Khanpur for the Mahi basin. The calibration and validation results for three basins are shown in Table \ref{Calibration and validation results}. The time-series of observed streamflow and VIC simulated monthly streamflow is given in the appendix (Figure \ref{FigA0}). We use Nash–Sutcliffe model efficiency coefficient (NSE)  {\cite{nash1970river}} and coefficient of determination (${R}^2$)  {\cite{nagelkerke1991note}} performance criteria, which are commonly used metrics in hydrological and climate impact studies. We use simulated monthly data and observed data for calculation of NSE and ${R}^2$ to analyze the performance of the VIC. We estimate streamflow using the bias and non-bias corrected data of each ensemble of MICE and MME separately using parameters obtained after calibration and validation. We also performed separate analyses for historical, near-term, mid-term and far-term data. We route the streamflow to the dam location to calculate potential hydropower.




\subsection{Potential Hydropower Estimation}
We calculate hydropower production for all the ensembles using the estimated streamflow. We calculate Developed Hydropower Potential (DHP), the maximum possible hydropower generated using the available water and existing hydroelectric facilities  {\cite{liu2016projected}}. First, we calculate the monthly releases using estimated streamflow and generic regulation rules described by  {\cite{hanasaki2006reservoir}}, which contain the effect of yearly and monthly variability. We calculate monthly release $R_m$ $(m^{3}/s)$ using Equation \ref{eq:1}, given by  {\cite{hanasaki2006reservoir}}, which can be used where irrigation is not considered.  {\cite{hanasaki2006reservoir}} has developed a reservoir operation scheme for global river routing models, which uses reservoir specifications, global runoff data sets, and water demand downstream. Here, we note that DHP does not represent the actual hydropower production as it does not consider socioeconomic factors such as water use variability, power consumption variation, and consumer behaviour. We assume that the reservoir is primarily used for hydropower production. The monthly release from the reservoir is generally not influenced by monthly inflow when storage capacity is large compared to the mean annual inflow. However, monthly release fluctuates if storage capacity and mean annual inflow are of the same order  {\cite{ali2018projected}}.  

\begin{equation} \label{eq:1}
  \begin{aligned}
    R_{\mathrm{m}}=\left\{\begin{array}{l}
    \left(\frac{c}{0.5}\right)^{2} k_{y} i_{\mathrm{a}}+\left[1-\left(\frac{c}{0.5}\right)^{2}\right] i_{\mathrm{m}},\:\:\:\:  0 < c < 0.5 \\
    k_{y} i_{\mathrm{a}} \:, \;\;\;\;\;\;\;\;\;\;\;\;\;\;\;\;\;\;\;\;\;\;\;\;\;\;\;\;\;\;\;\;\;\;\;\;\;\;\;\;\:\:\:\:\:\: c > 0.5
  \end{array}\right.
  \end{aligned}
\end{equation}

Where c is the ratio of maximum storage capacity to mean total annual inflow,  $c = C/Ia$. C is the maximum storage capacity of the reservoir ($m^{3}$),  $I_{a}$ is the average total annual inflow ($m^{3}/year$). $k_{y}$ is the release coefficient that considers water stored in the reservoir at the beginning of the operational year  {\cite{hanasaki2007integrated, ali2018projected}}. It is calculated as $k_{y}=S_{beg}/{\alpha}\times{C}$, $S_{beg}$ is the storage available at the beginning of the operational year, ${\alpha}$  is an empirical coefficient suggested based on sensitivity analysis. In this study, ${\alpha}$ value considered is 0.85 as suggested by  {\cite{hanasaki2007integrated}}, $i_{m}$ is monthly inflow  ($m^{3}/s$) and  $i_{a}$ is mean annual inflow  ($m^{3}/s$). Two constraints are considered for the analysis. The first is that monthly release should not be less than 10 \% of the monthly inflow to the reservoir. The second is that at any point, the storage in the reservoir should not be greater than storage capacity, and a minimum of 10 \% of storage capacity is maintained for flood control and maintaining minimum water levels. We calculated the storage using the continuity equation and compared calculated storage with observed storage (Billion Cubic Meters: BCM), which was downloaded from India-WRIS to ensure the method's accuracy. We observe that the calculated storage values are slightly higher than those observed, as we have not considered irrigation demands and other demands for the analysis. Then we calculate the height using the storage height relationship. DHP can be calculated based on monthly release, storage capacity, and Installed Hydraulic Capacity (IHC)  {\cite{liu2016projected}}. The equation to calculate DHP is given by Equation \ref{eq:2}, which ensure that it will not exceed IHC.  

\begin{equation} \label{eq:2}
\mathrm{DHP}=\min \left(\mathrm{R}_{\mathrm{m}} \times \mathrm{h} \times \mathrm{g}, \mathrm{IHC}\right)
\end{equation}
where $R_m$ is monthly release, $h$ is a hydraulic head (m) and $g$ is an acceleration of gravity ($m/s^2$). We analyze the uncertainty resulting from internal variability and model uncertainty in estimating hydropower generated for four different periods: historical, near-term, mid-term, and end-term. The monthly power production and uncertainties resulting from internal variability and model uncertainties are analyzed for each period. We also analyze percentage change with respect to the historical period of streamflow and hydropower production to analyze uncertainties and their translation from streamflow to hydropower production.
 
\begin{table}
\caption{Calibration and validation results}
\label{Calibration and validation results}
\centering
\begin{tabular}{|p{1.8cm}|p{2.4cm}|p{0.2cm}p{0.2cm}p{1.8cm}|p{0.3cm}p{0.3cm}p{1.8cm}|}
\hline
 \multirow{2}{*}{\textbf{Basin}}   & \multirow{2}{*}{\textbf{Station name}} & \multicolumn{3}{c|}{\textbf{Calibration}} & \multicolumn{3}{c|}{\textbf{Validation}}  \\ \cline{3-8}  &  & \multicolumn{1}{c|}{\textbf{NSE}} & \multicolumn{1}{c|}{\textbf{R-square}} & \textbf{Period}    & \multicolumn{1}{c|}{\textbf{NSE}} & \multicolumn{1}{c|}{\textbf{R-square}} & \textbf{Period}    \\ \hline
 \multirow{3}{*}{\textbf{Narmada}} & Handia                                 & \multicolumn{1}{c|}{0.90}         & \multicolumn{1}{c|}{0.92}              & Jan-1985-Dec-1999  & \multicolumn{1}{c|}{0.86}         & \multicolumn{1}{c|}{0.94}              & Jan-2000-Dec-2014  \\ \cline{2-8} 
                                  & Mandleshwar                     & \multicolumn{1}{c|}{0.94}         & \multicolumn{1}{c|}{0.94}              & Jan-1985-Dec-1999  & \multicolumn{1}{c|}{0.71}         & \multicolumn{1}{c|}{0.86}              & Jan-2000-Dec-2014  \\ \cline{2-8} 
                                  & Sandia                             & \multicolumn{1}{c|}{0.86}         & \multicolumn{1}{c|}{0.88}              & Jan-1985-Dec-1999  & \multicolumn{1}{c|}{0.86}         & \multicolumn{1}{c|}{0.89}              & Jan-2000-Dec-2014  \\ \hline
 \multirow{2}{*}{\textbf{Tapi}}    & Sarangkheda                            & \multicolumn{1}{c|}{0.90}         & \multicolumn{1}{c|}{0.92}              & Jan-1978-Dec-1986  & \multicolumn{1}{c|}{0.88}         & \multicolumn{1}{c|}{0.96}              & Jan-1987-Apr-1996  \\ \cline{2-8} 
                                  & Burhanpur                                 & \multicolumn{1}{c|}{0.82}         & \multicolumn{1}{c|}{0.89}              & Jan-1973-Dec-1980  & \multicolumn{1}{c|}{0.84}         & \multicolumn{1}{c|}{0.89}              & Jan-1981-Dec-1987  \\ \hline
\multirow{2}{*}{\textbf{Mahi}}    & Mataji                              & \multicolumn{1}{c|}{0.81}         & \multicolumn{1}{c|}{0.83}              & June-2005-May-2011 & \multicolumn{1}{c|}{0.78}         & \multicolumn{1}{c|}{0.80}              & June-2011-May-2017 \\ \cline{2-8} 
                                  & Khanpur                          & \multicolumn{1}{c|}{0.77}         & \multicolumn{1}{c|}{0.85}              & Jan-1985-Dec-1994  & \multicolumn{1}{c|}{0.62}         & \multicolumn{1}{c|}{0.83}              & Jan-1995-Dec-2004  \\ \hline
\end{tabular}
\end{table}

\section{Results and Discussions}
The calibration and validation results for Narmada, Tapi and Mahi basins are shown in Table \ref{Calibration and validation results}. We simulate streamflow using the observed precipitation, $T_{max}$, $T_{min}$, and wind data in the VIC model and compared it with observed data. We consider  0.8 as threshold for performance criteria (NSE and ${R}^2$) as suggested by previous studies ( {\cite{huang2017evaluation}}. Table \ref{Calibration and validation results} shows that values of NSE and ${R}^2$, which are higher than the threshold for all gauge stations except Khanpur gauge. We analyzed the time series of simulated streamflow and compared it with observed data, which shows better agreement with each other (Figure \ref{FigA0}). However, simulated streamflow overestimates observed streamflow for the Mahi basin's Khanpur gauge. It shows NSE values as 0.77 and 0.62 for the calibration and validation period, respectively (Table \ref{Calibration and validation results}), which may be due to its location (downstream of the dam Figure \ref{Fig1}). We have not incorporated the reservoir module into the VIC model to simulate the streamflow  {\cite{yun2020impacts}}.

\begin{figure}
\begin{center}
    \fbox{\includegraphics[width=0.95\textwidth,keepaspectratio]{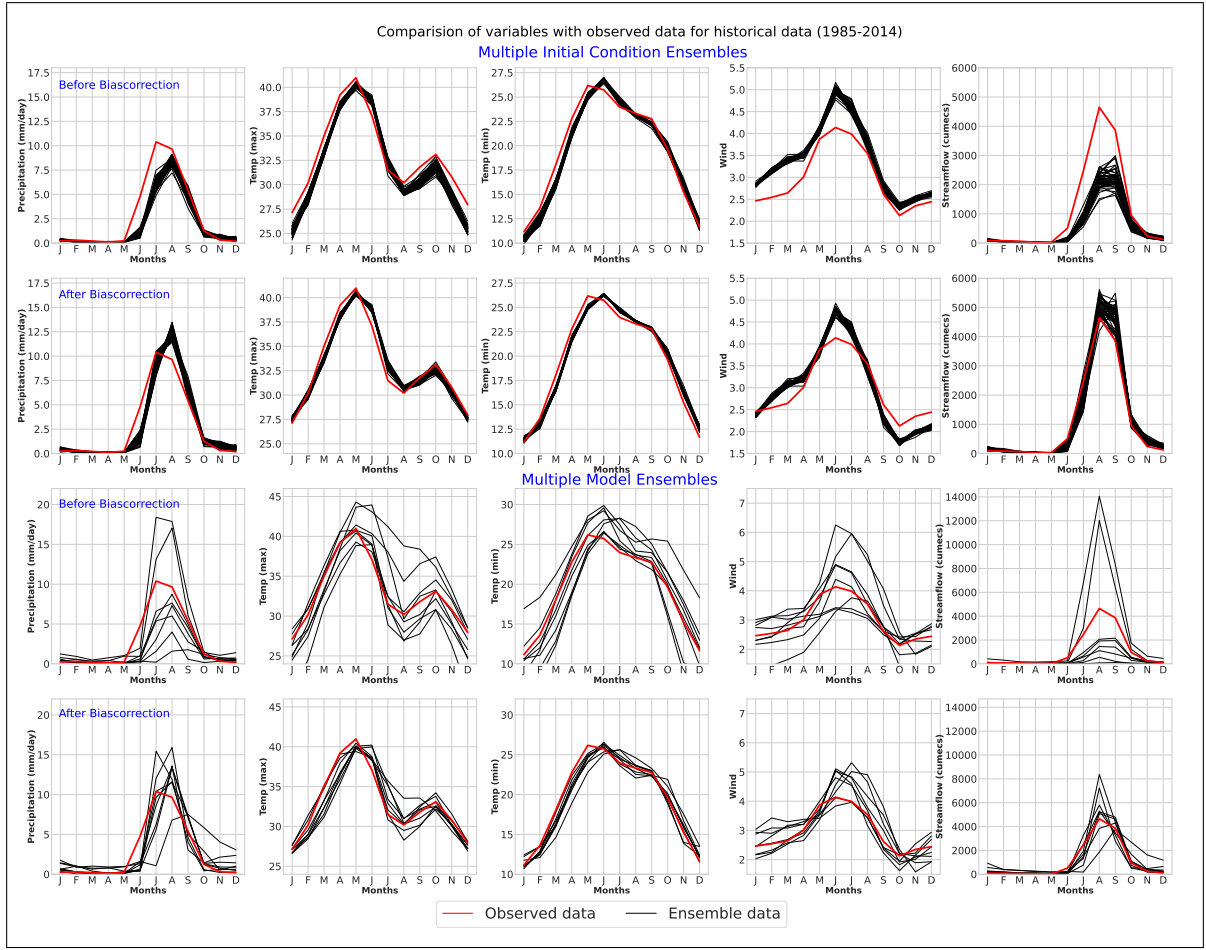}}
    \caption{\textbf{Effect of Bias correction}: Bias correction does not preserve the internal variability in estimating the streamflow. The first and third row shows the monthly precipitation, $T_{max}$, $T_{min}$, wind, and estimated streamflow using the non-bias corrected data, respectively. The second and fourth row shows the monthly variation of the same variables using the bias-corrected data. Each black line represents the monthly mean average of the historical period (1985-2014) over an area from the individual ensembles (MICE/MME) considered in this study. The red line shows observed monthly mean values averaged over the basin up to the dam location for precipitation, $T_{max}$, $T_{min}$, wind, and it shows streamflow at dam location. A similar result for Kadana and Ukai is shown in the appendix (Figure \ref{FigA1}). Analysis for near-term, mid-term and far-term is also shown in the appendix (Figure \ref{FigA2}).}
    \label{Fig3}
    \end{center}
\end{figure}

We analyze the effect of bias correction of meteorological variables such as P, $T_{max}$, $T_{min}$, W and its impact on streamflow estimation for each ensemble from MICE and MME. We compare monthly mean meteorological variables and estimated streamflow of each ensemble with the observed data. Figure \ref{Fig3} shows variations of meteorological variables of the Narmada basin and estimated streamflow at Sardar Sarovar dam for the historical data. Non-bias corrected monthly precipitation from MICE shows lower precipitation than observed for monsoon (June - September). We observe that bias-corrected precipitation tries to match for July and increase precipitation values for August and September with respect to observed data. For $T_{max}$ and  $T_{min}$, bias correction does not significantly improve the performance. Mean monthly non-bias corrected wind from each ensemble of MICE shows higher values than observed wind data for all months. The difference between bias-corrected mean wind data and observed data reduces after bias correction. Still, the difference between ensemble data and observed data shows that quantile-quantile mapping does not remove bias effectively for wind data. The streamflow estimated using meteorological non-bias corrected data underestimates compared to observed data, while for bias-corrected data shows better agreement with the observed data (The last column of Figure \ref{Fig3}). However, we observe that streamflow estimated using bias-corrected data slightly overestimates compared to observed data for August and September. Overall, mean monthly data from each ensemble of MICE shows better agreement with each other, indicating lower ICV as shown in the first and second row of Figure \ref{Fig3}. We also observe that bias correction further reduces ICV, indicating that it cannot preserve the ICV, which is considered irreducible. 

The mean monthly variations of P, $T_{max}$, $T_{min}$, wind and estimated streamflow of ensembles from MME and its comparison with observed data shows that model uncertainty is higher than ICV. Mean monthly precipitation exhibit higher model uncertainty for monsoon than $T_{max}$, $T_{min}$ and wind. This model uncertainty also translates to streamflow and shows the higher model uncertainty in estimating the streamflow. For example, two ensembles overestimate streamflow than observed, while other ensembles underestimate for non-bias corrected data, especially for monsoon. Bias corrected data of MME shows an improvement for all the variables and tries to match with observed data. However, we observe high model uncertainties even after bias correction as quantile-quantile mapping bias correction does not seem able to handle high variability and seasonality. We performed a similar analysis for Kadana and Ukai (Figure \ref{FigA1}). Meteorological variables show similar behaviour for Tapi and Mahi basins as the Narmada basin. There is a significant difference between streamflow estimated using bias and non-bias corrected data for Kadana dam, which is not observed for Ukai dam. This difference between streamflow estimated using the bias and non-bias corrected data is observed for other periods, such as near-term, mid-term and far-term, especially for monsoon months. (Figure \ref{FigA2}). 

\begin{figure}
\begin{center}
    \fbox{\includegraphics[width=0.95\textwidth,keepaspectratio]{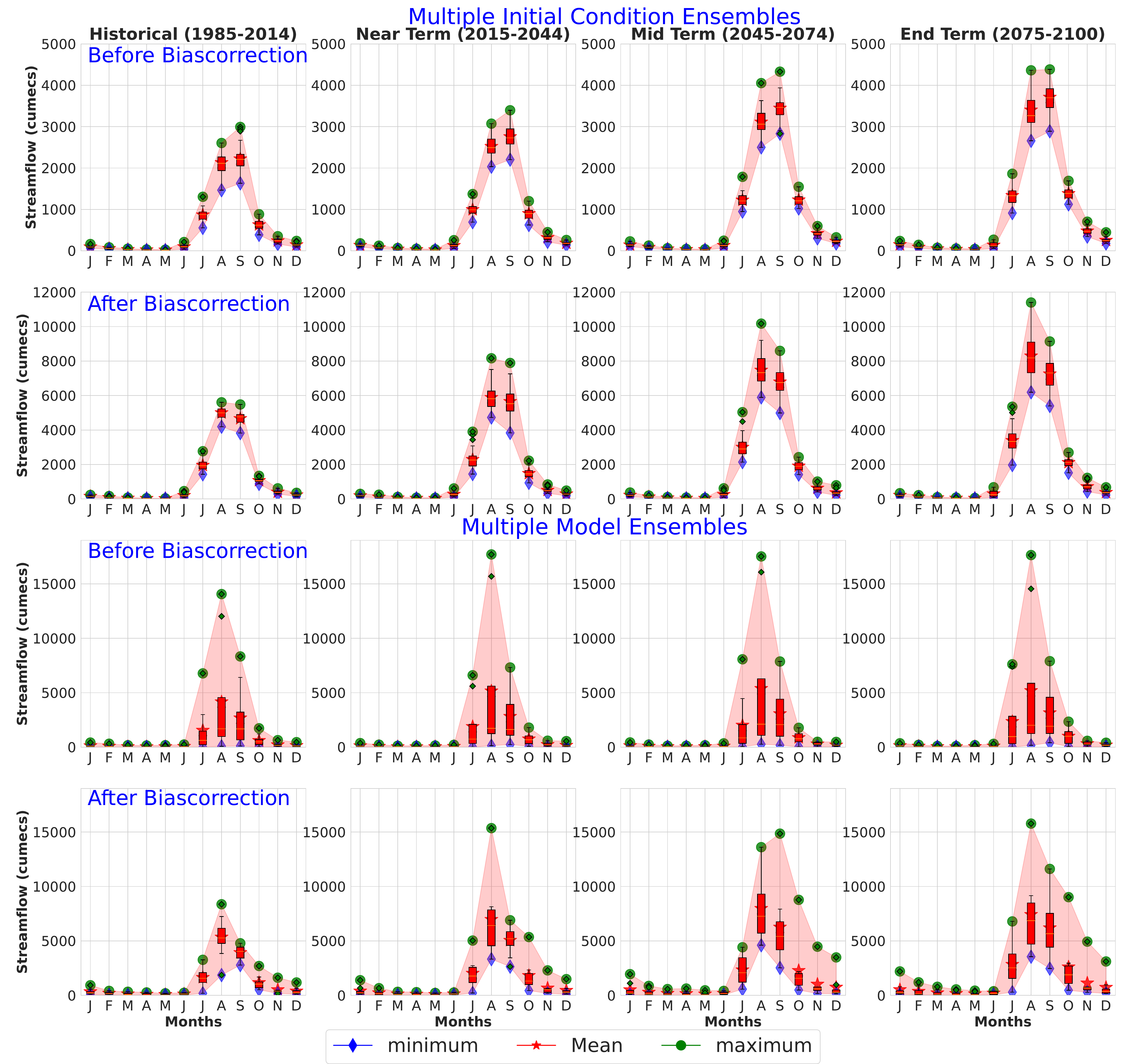}}
    \caption{\textbf{Uncertainties for streamflow}: IQR for estimated streamflow for historical (1985-2014), near-term (2015-2044), mid-term (2045-2074), and far-term (2075-2100) using bias and non-bias corrected data of Sardar Sarovar. The first and second row shows the uncertainties due to ICV in estimating streamflow, estimated using non-bias and bias-corrected data, respectively, using Multiple Initial Condition Ensembles. Similarly, the third and fourth row shows model uncertainty, obtained using the multiple model ensembles. A similar result for Kadana and Ukai is shown in the appendix (Figure \ref{FigA3}).}
    \label{Fig4}
    \end{center}
\end{figure}

To analyze the role of ICV and model uncertainty in estimating the streamflow, we consider different periods such as historical (1985-2014), near-term (2015-2044), mid-term (2045-2074), and far-term (2075-2100) (Figure \ref{Fig4}). We consider interquartile range (IQR - the difference between the 75th and 25th percentile) as the measure for uncertainty in estimating the streamflow within ensembles (MICE/MME) {\cite{upadhyay2021depth}}. Figure \ref{Fig4} shows IQR in estimating streamflow at Sardar Sarovar dam for all four periods. We observe lower internal variability and model uncertainty for the streamflow estimated using the bias-corrected data than non-bias corrected data for historical periods, indicating the effect of bias correction. We also note a significant difference between the mean monthly streamflow estimated using bias and non-bias corrected data. For example, mean monthly stream flows are approximately 2000 cumecs and 5000 cumecs when estimated using August's historical non-bias and bias-corrected data. This difference further increases from near-term to far-term, as shown in the appendix (Figure \ref{FigA2}). However, Mean streamflow and uncertainty due to ICV increases from near-term to far-term for MICE. The third and fourth row of Figure \ref{Fig4} shows that model uncertainty is higher than ICV and increase in mean streamflow from near-term to far-term for MME. Higher model uncertainty in estimating streamflow is observed for monsoon months (Figure \ref{Fig4}). Model uncertainty reduces using bias-corrected data, especially for a historical period. We observe higher mean streamflow, ICV and model uncertainty only for monsoon and post-monsoon (October-December), which agrees with other studies {(\cite{shah2018climate}, \cite{nilawar2019impacts})}. Similarly, we analyze the results for Kadana and Ukai dams as given in the appendix (Figure \ref{FigA3}), which also shows similar patterns of streamflow uncertainties. Here, We highlight the uncertainty induced due to ICV increases towards the far-term irrespective of bias correction. Thus, it is important to internalize the effect of internal variability for planning and to manage water resources systems.

\begin{figure}
\begin{center}
    \fbox{\includegraphics[width=0.95\textwidth,keepaspectratio]{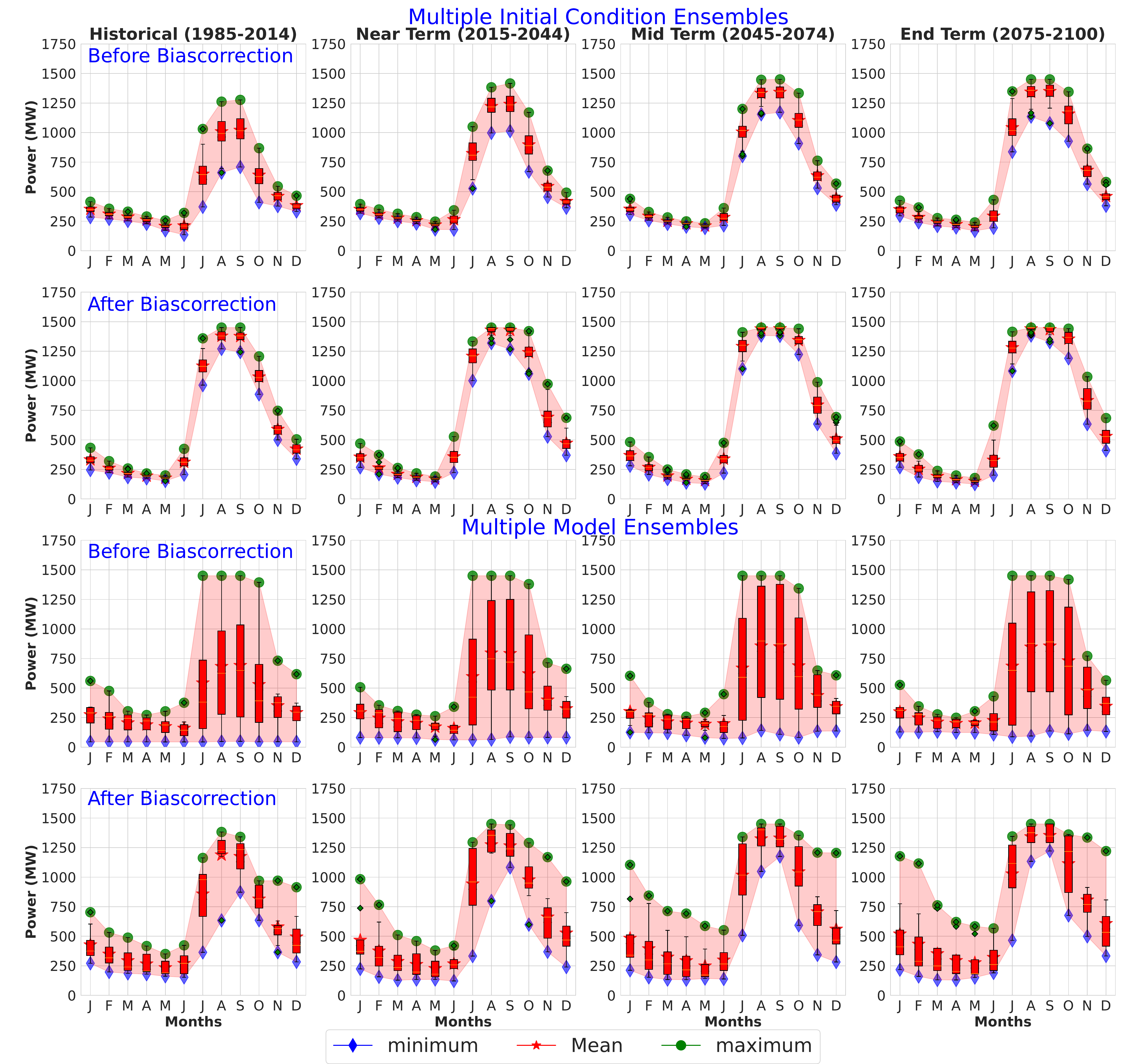}}
    \caption{\textbf{Uncertainties for power}: IQR for estimated power for historical (1985-2014), near-term (2015-2044), mid-term (2045-2074), and far-term (2075-2100) using bias and non-bias corrected data for Sardar Sarovar. The first and second row shows the ICV in estimating the potential hydropower, estimated using non-bias and bias-corrected data using Multiple Initial Condition Ensembles. Similarly, the third and fourth row shows the model uncertainty obtained using the multiple model ensembles. A similar result for Kadana and Ukai is shown in the appendix (Figure \ref{FigA4}).}
    \label{Fig5}
    \end{center}
\end{figure}

\begin{figure}
\begin{center}
    \fbox{\includegraphics[width=0.95\textwidth,keepaspectratio]{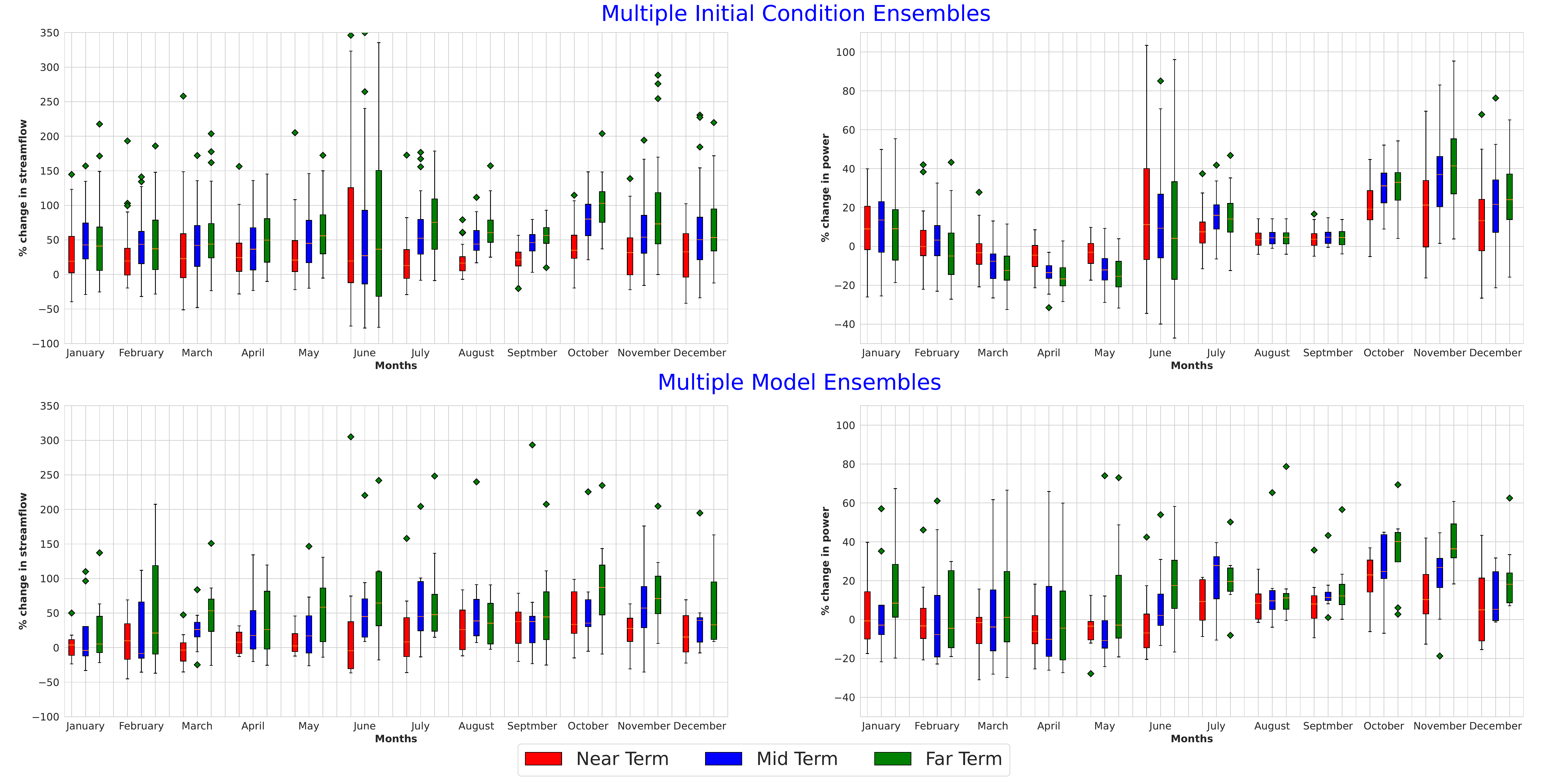}}
    \caption{\textbf{Percentage change of streamflow and hydropower}: Percentage change in streamflow is not reflected in percentage change in hydropower. The figure shows the percentage change of mean monthly streamflow and hydropower to historical data using bias-corrected data for Sardar Sarovar. Uncertainty due to internal variability and model uncertainty in estimating the streamflow and hydropower are shown in the first and second row, respectively. A similar result for Kadana and Ukai is shown in the appendix (Figure \ref{FigA5}).}
    \label{Fig6}
    \end{center}
\end{figure}

Uncertainty analysis for potential hydropower shows that ICV and model uncertainty in estimating the mean monthly hydropower production plays a significant role for all months. Figure \ref{Fig5} shows IQR for mean monthly potential hydropower estimated using the historical, near-term, mid-term and far-term streamflow data for Sardar Sarovar dam. Uncertainties in estimating the hydropower are also lower for bias-corrected data as we observe for streamflow. The reduction in ICV shows that bias correction does not preserve the internal variability. A significant increase in mean hydropower production is observed between June and July for mid and end-term for non-bias corrected data of MICE, while historical data shows the smooth transition. However, we observe a similar increase for all periods using bias-corrected data as it represents the start of the monsoon. MICE data capture the phenomenon due to starting of the monsoon. We observe uniform and lower internal variability throughout all months than model uncertainty. There is no significant increase in uncertainty towards the end of 21 st century (end-term) as calculation depends on the hydropower plant's installed capacity and storage capacity. However, we observe that the lower bound of IQR increases towards the end-term, while the upper bound remains almost constant for monsoon and post-monsoon. The third and fourth row of Figure \ref{Fig5} shows the model uncertainty in estimating the hydropower production. We observe higher model uncertainty in estimating hydropower production, especially for monsoon for non-bias corrected data, while bias-corrected data shows higher model uncertainty for non-monsoon. Similarly, model uncertainty increases towards the end-term for non-monsoon months, which we do not observe for estimated streamflow. We also performed a similar analysis for Kadana and Ukai dams as given in the appendix (Figure \ref{FigA4}), which also shows a similar pattern as the Sardar Sarovar dam. However, uniform hydropower production, model uncertainty and ICV in estimating hydropower are almost constant throughout the year for Ukai. Overall, we can say that the uncertainty bounds for hydropower production are not similar to the streamflow. Hydropower estimation shows a contribution of internal variability and model uncertainty throughout the year than streamflow uncertainty.

For further analysis, we performed the percentage change analysis to compare the role of ICV and model uncertainty in estimating streamflow and hydropower. Figure \ref{Fig6} shows the percentage change of mean monthly streamflow and hydropower of near-term, mid-term and end-term with respect to historical bias-corrected data for Sardar Sarovar. Mean percentage change shows the increase of streamflow for all months. However, this increase in streamflow is not reflected in hydropower production. For example, mean hydropower production decrease towards the far-term for February to May. This decrease is more prominent for MICE than MME. On the contrary, potential hydropower production for monsoon and post-monsoon shows an increase in hydropower production. We observe significant variability in percentage change obtained using MICE for June, which indicates the starting of the monsoon such as percentage change in ICV ranges between $-40$ \% to 150 \% for streamflow and $- 20$ \% to 40 \% for hydropower (First row of Figure \ref{Fig6}). In comparison, model uncertainty does not show this significant change.

Mean percentage change in model uncertainty in streamflow estimation and variability increase from near-term to end-term. Similarly, percentage changes analysis of streamflow and hydropower is also performed for Kadana and Ukai as given in the appendix (Figure \ref{FigA5}). For Kadana, percentage change clearly shows an increase in mean and uncertainty in estimating streamflow and hydropower production towards the end of the century (far-term) (Figure \ref{FigA5} (a)). The percentage change variability in estimated streamflow and hydropower due to ICV is even higher for June, as observed for Sardar Sarovar. However, a similar trend is not observed for the Ukai dam. Figure \ref{FigA5} (b) shows higher percentage change and its variability in estimating streamflow for Ukai dam, especially for ICV. However, similar variability and increase in mean percentage change are not observed for potential hydropower. Despite an increase in streamflow, a decrease in hydropower production is observed for February to April. Overall, The percentage change for streamflow is approximate $-48$ \% to 240 \%, while for hydropower is approximately $-20 $ \% to 70 \% change for all cases, which indicates that increasing streamflow does not increase hydropower production significantly. 

\section{Conclusions}
Hydropower production can help meet the increasing energy demands due to urbanization and the increasing population in India. Hydropower, the extensively used renewable energy sensitive to streamflow change, is characterized by various uncertainties. The improved understanding and better prediction of hydropower empower the stakeholders for planning and decision-making, such as to estimate the potential energy production and determine design parameters such as water level, minimum operation level and installation capacity of hydropower based on future projections. However, the crucial role of internal climate variability on hydropower production in India is still unexplored. The high variability in precipitation, $T_{max}$, $T_{min}$ and wind, which directly or indirectly affects the streamflow and hydropower production and acknowledging the importance of internal variability, motivated us to explore further the role of internal climate variability in projections of future hydropower production. This study analyzes the role of internal climate variability, which is irreducible uncertainty in estimating streamflow and hydropower production. We estimate streamflow using VIC hydrological model. This study analyzes the role of internal variability and model uncertainty in estimating the streamflow and its translation for hydropower estimation. We have also analyzed the effect of bias correction by comparing the results of bias and non-bias corrected data. The result shows that bias correction does not preserve the internal variability in streamflow estimating, which further impacts the potential hydropower estimation. The other findings from this study are as follows. Model uncertainty contributes more to total uncertainty than ICV in estimating the streamflow and potential hydropower for monsoon. However, we observe that role of ICV is increasing towards the far-term. The estimated mean streamflow increases during monsoon and post-monsoon only. For non-monsoon, both uncertainties play a significant role in hydropower estimation, while it is not evident from estimated streamflow. The uncertainty resulting from ICV and model uncertainty increases from near-term to far-term for estimated streamflow, while we have not observed a similar increase for hydropower production. However, the lower bound of IQR increases towards the far-term (2075-2100) compared to the historical data for hydropower production. Mean potential hydropower shows the decrease towards the far-term, especially for February to May, despite increasing uncertainty for estimated streamflow. This decrease in hydropower is more prominent for MICE than MME. The higher percentage change in ICV is observed for streamflow ($-40$ \% to 150 \%) than hydropower ($-20$ \% to 40 \%) for June. In comparison, model uncertainty does not show this significant change.

This paper uses quantile mapping for bias correction for individual ensembles of MICE and MME. Recently, {\cite{ayar2021ensemble}} have developed ensemble bias correction method to preserve internal variability. Further study may consider using the ensemble bias correction approach to understand the role of ICV better. The present study considers fixed duration data for investigation using a historical period (1985-2014) and future periods such as 2015-2044 (near-term), 2045-2074 (mid-term), and 2075-2100 (far-term). The future scope includes the trend analysis for hydropower production. We use VIC hydrological model for the estimation of streamflow. We have not considered hydrological model uncertainty for the analysis, which may play a significant role in estimating the streamflow. Thus, other studies may consider the combined effect of climate models, hydrological model uncertainty and internal variability in estimating hydropower. In this study, we have calculated the monthly releases using generic regulation rules described by  {\cite{hanasaki2006reservoir}}. Further study may consider applying different reservoir operating policies for calculating the monthly release by considering the irrigation and municipal water demand under different demand scenarios. We use 50 multiple initial condition ensembles from only one model (EC-Earth3) to quantify the internal variability by assuming that it captures all the variability and represents the internal climate variability. There are multiple Single Model Initial Condition Large Ensembles (SMILEs) that are now available, which may provide new insights for analyzing the role of internal climate variability  {\cite{lehner2020partitioning, wood2020analyzing, deser2020insights}}. 

\section*{Acknowledgments}
This work is supported by DST-SERB Startup Research grant (SRG/2019/000833) and the DST-SCCC-NMSKCC project (RES/DST/EH/P0137/1920/0005), work for Establishing/Strengthening the State Climate Change Centre/Cell under NMSKCC (SCCC-NMSKCC) in the State of Gujarat, India. High-resolution observations over three basins were distributed by Indian Meteorological Department located at Pune. Open-source datasets are obtained from PCMDI, NCAR and India WRIS. The multiple initial condition ensembles of the CMIP6-EC-Earth data is downloaded from Earth System Grid Federation \url{(https://esgf-data.dkrz.de/projects/esgf-dkrz/)}. The multiple models of CMIP6 are downloaded from World Climate Research Program Coupled Model Intercomparison Project \url{(https://esgf-node.llnl.gov/projects/cmip6/)}. The authors thank Prof. Srikrishnan Subramanian, Pravin, Angana and Raviraj for comments on the manuscript. The authors would also like to thank Prof. Vimal Misra, Amar Deep Tiwari, and other members of the Water and Climate Lab of IIT Gandhinagar, India, for sharing the wind observed data, soil and vegetation parameters and providing help to set up the VIC hydrological model.

\section*{Author Contributions}
Udit Bhatia,  Sudhanshu Dixit and Divya Upadhyay designed the experiments.  Sudhanshu Dixit and Divya Upadhyay performed the experiments and analyzed the data. Divya Upadhyay, Udit Bhatia and Sudhanshu Dixit wrote the manuscript.
\clearpage
\bibliographystyle{unsrt}  
\bibliography{references} 

\newpage
\section*{Appendix}
\appendix

\begin{table}
\renewcommand\thetable{A1}
\begin{center}
\caption{Details of Powerplants}
\label{Details of powerplant}

\begin{tabular}{ |p{4cm}|p{2.5cm}|p{3cm}|p{3cm}| }
\hline
 Dam & Sardar Sarovar & Ukai & Kadana \\
 \hline
Total installed capacity (MW) & 1450 & 305 & 240\\
  \hline
Latitude & \ang{21.86} N & \ang{21.35} N & \ang{23.29}N\\
\hline
Longitude & \ang{73.91} E & \ang{73.80} E & \ang{73.82} E\\
\hline
State & Gujarat & Gujarat & Gujarat\\
\hline
River / Basin & Narmada & Tapi & Mahi\\
\hline 
Catchment area of Dam (sq. km) & 88,000 & 62,255 & 25,520\\
\hline

Type of Dam & Gravity & Earthen,Masonry & Earthen,Masonry\\
\hline
Full reservoir level (m) & 138.68 & 105.16 & 127.7\\
\hline
 \end{tabular}
\end{center}
\end{table}

\begin{table}
\renewcommand\thetable{A2}
\caption{The list of CMIP6 Models used in this study}
\label{List of CMIP6 data}
\begin{center}
\begin{tabular}{ | m{0.5cm} | m{7cm}| m{2.7cm} | m{2.3cm} |}
 \hline
\textbf{No} & \textbf{Modeling Group} & \textbf{Model Name} & \textbf{Resolution (Degrees)}    \\
\hline
1 & Australian Community Climate and Earth System Simulator, Australia & ACCESS$-$CM2 &\ang{1.9} X \ang{1.3} \\ \hline

2 &  Canadian Centre for Climate Modelling and Analysis, Canada & CanESM5 & \ang{2.8} X \ang{2.8}\\ \hline

3 &  Consortium of European research institution and researchers, Europe & EC$-$EARTH3 & \ang{0.7} X \ang{0.7}\\ \hline

4 &  Geophysical Fluid Dynamics Laboratory (GFDL), USA & GFDL$-$ESM4 & \ang{1.25} X \ang{1}\\ \hline

5 & Marchuk Institute of Numerical Mathematics, Russia & INM$-$CM4$-$8 & \ang{2} X \ang{1.5}\\ \hline

6 &  Marchuk Institute of Numerical Mathematics, Russia & INM$-$CM5$-$0 & \ang{2} X \ang{1.5}\\ \hline

7 &  Meteorological Research Institute, Japan & MRI$-$ESM2$-$0 & \ang{1.1} X \ang{1.1}\\ \hline

8 &  Norwegian Climate Centre/Norway & NorESM2$-$LM & \ang{1.875} X \ang{2.5}\\ \hline
\end{tabular}
\end{center}
\end{table}

\begin{figure}
\renewcommand{\thefigure}{A1}
\centering  
\subfigure[Narmada Basin]{\includegraphics[width=0.45\linewidth]{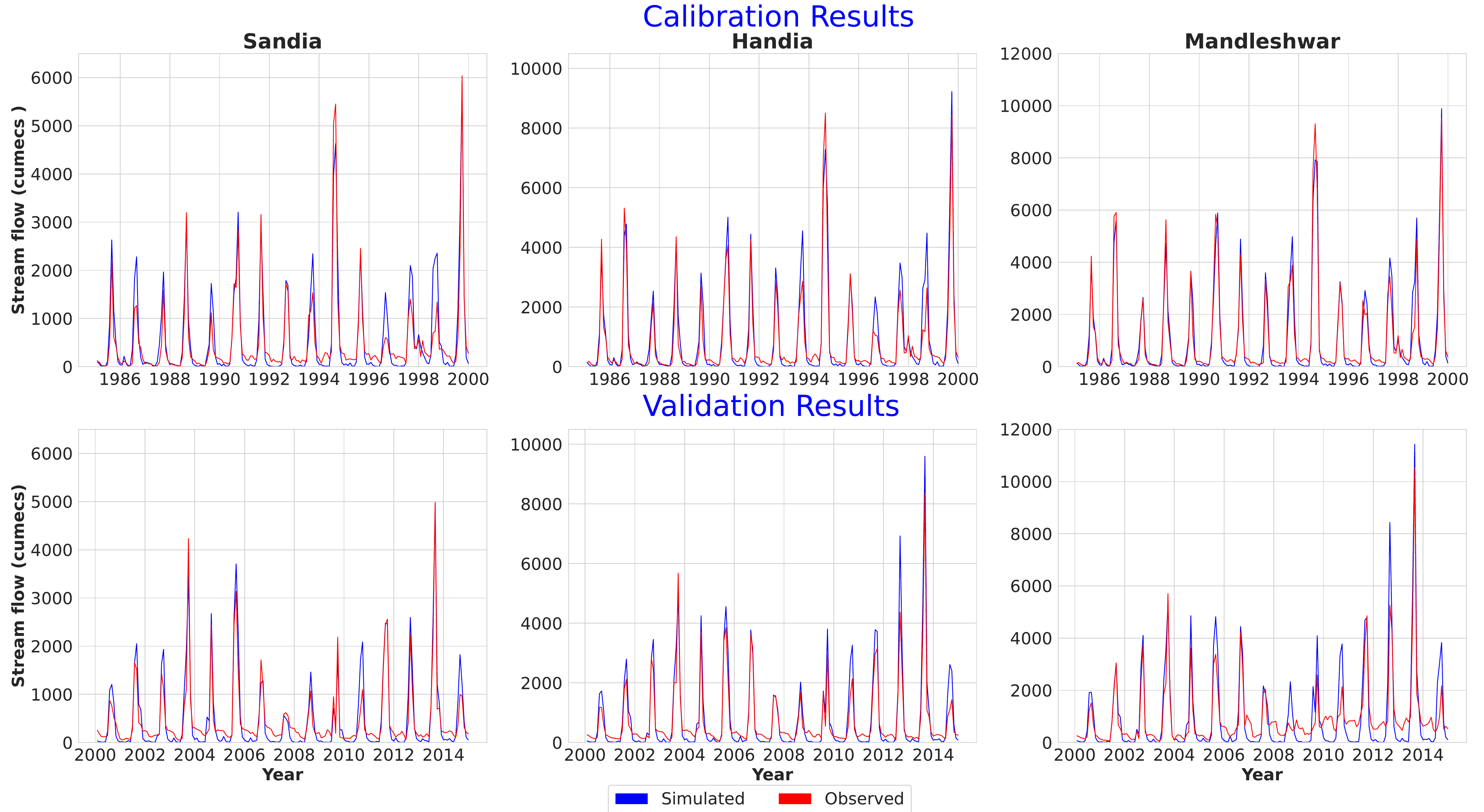}}
\subfigure[Mahi Basin]{\includegraphics[width=0.45\linewidth]{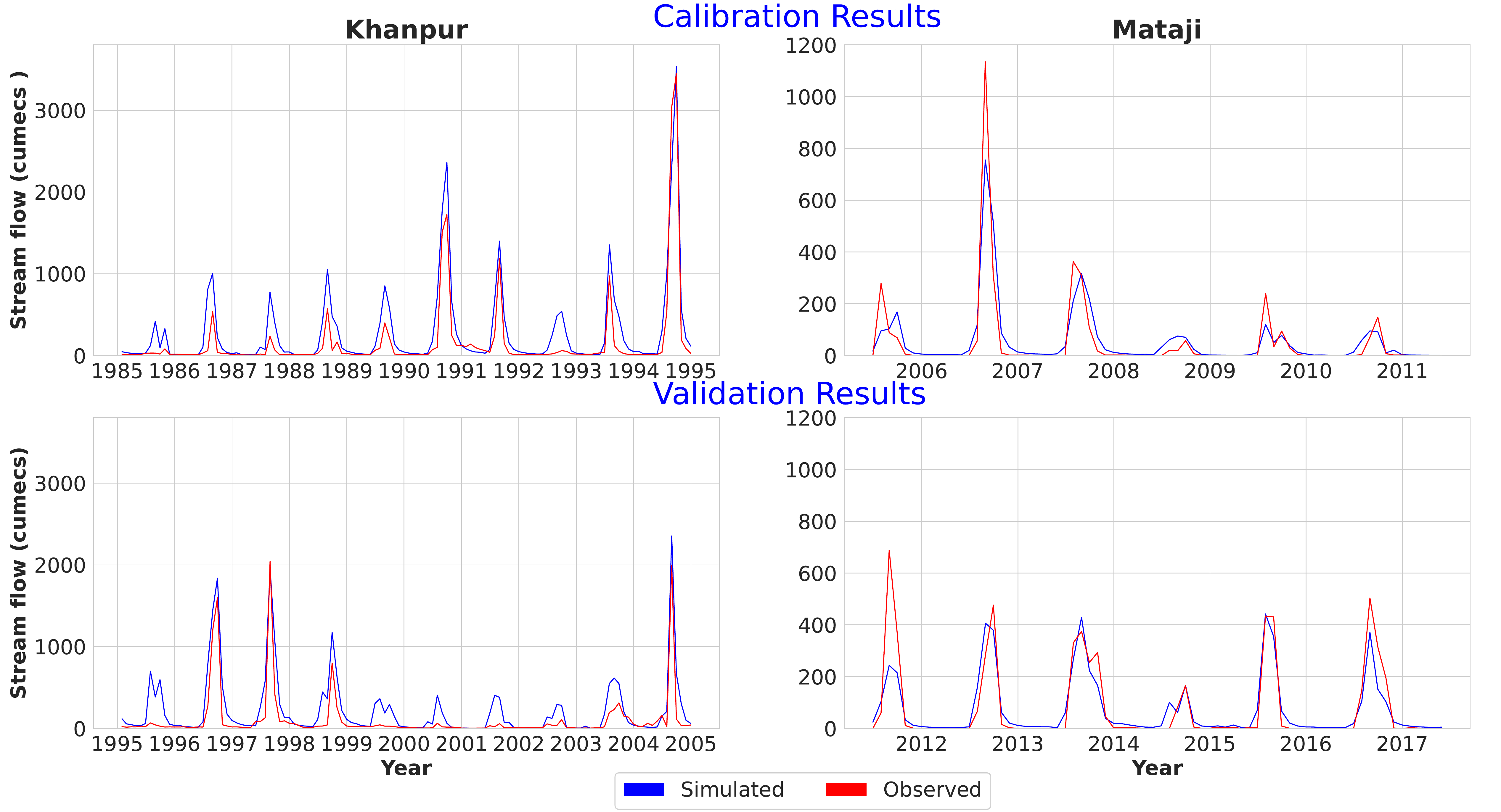}}
\subfigure[Tapi Basin]{\includegraphics[width=0.45\linewidth]{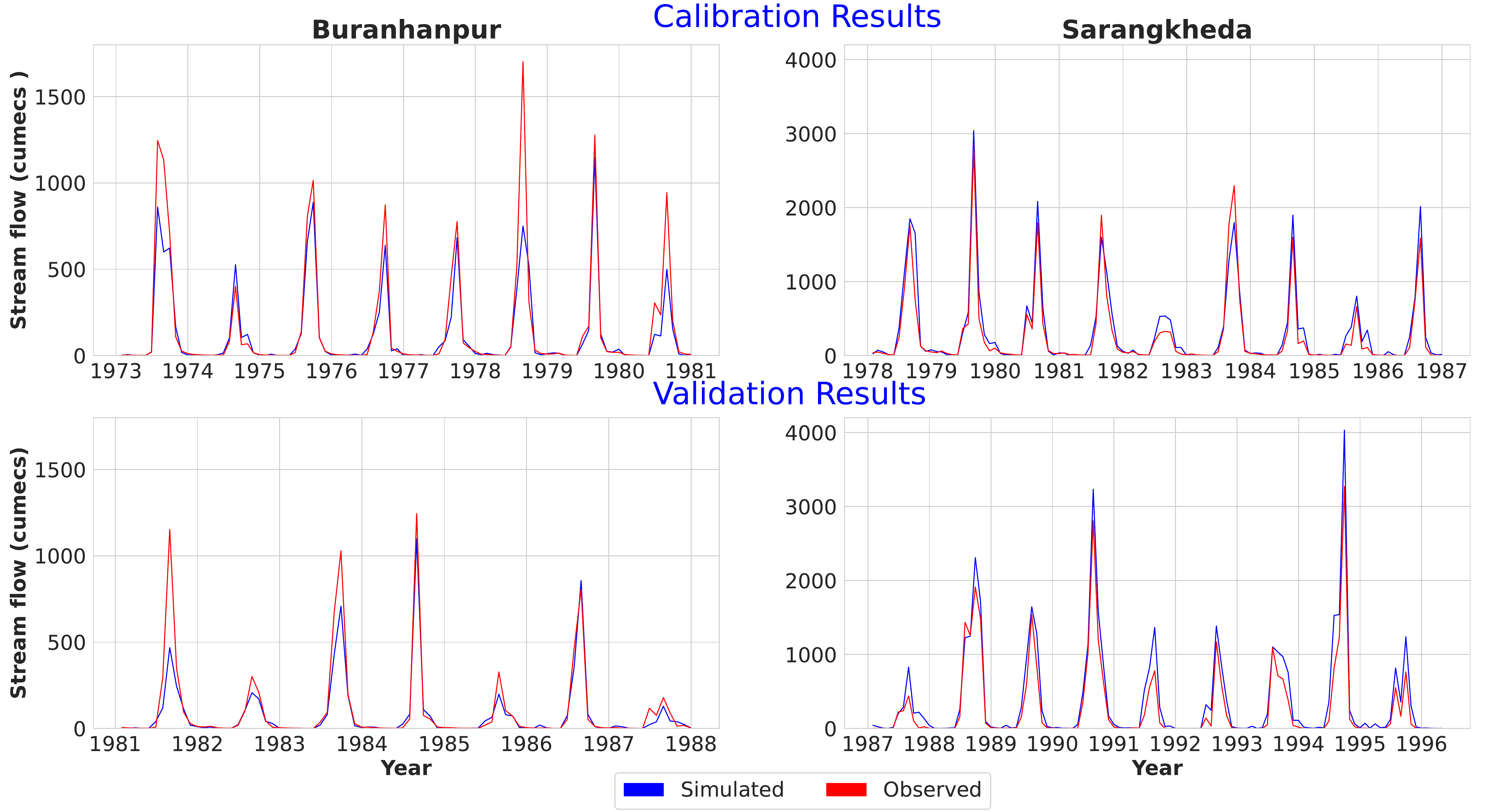}}
\caption{\textbf{Calibration and validation results}: Blue line shows the VIC simulated monthly streamflow at different gauges, and the red line shows observed monthly streamflow at different gauges}
\label{FigA0}
\end{figure}

\begin{figure}
\renewcommand{\thefigure}{A2}
\centering  
\subfigure[kadana]{\includegraphics[width=0.45\linewidth]{Figure_A2_1.pdf}}
\subfigure[Ukai]{\includegraphics[width=0.45\linewidth]{Figure_A2_2.pdf}}
\caption{Figure shows similar result as shown in figure 3 for Kadana and Ukai}
\label{FigA1}
\end{figure}

\begin{figure}
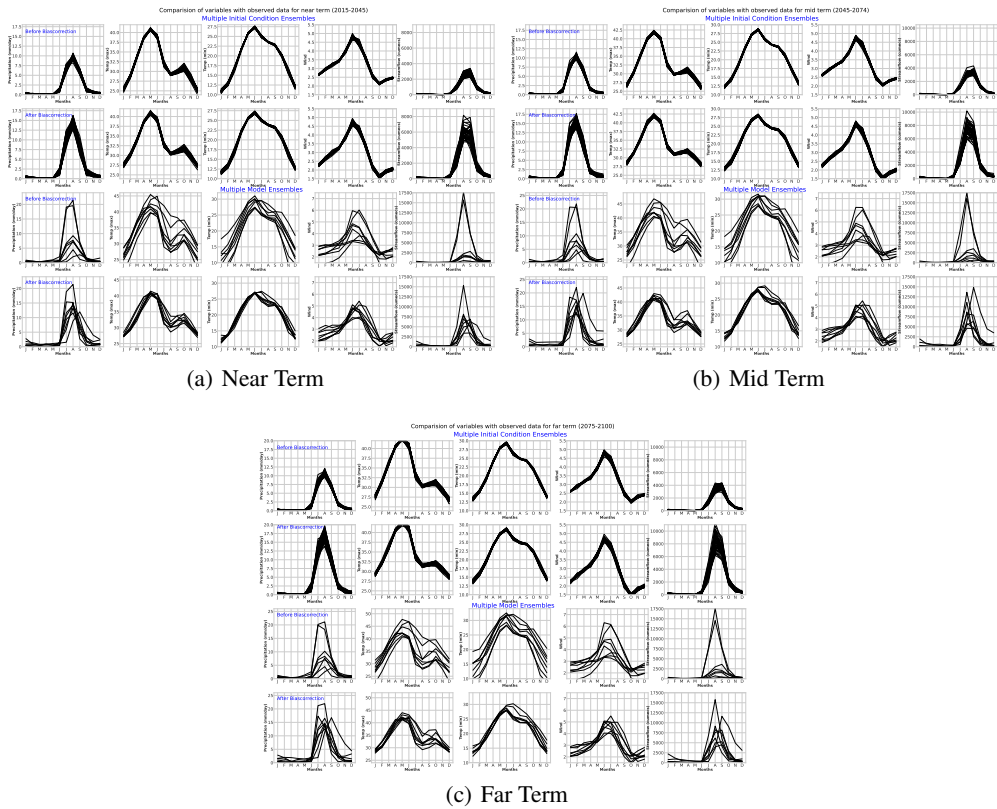

\renewcommand{\thefigure}{A3}
\centering  
\subfigure[Near Term]{\includegraphics[width=0.4\linewidth]{Figure_A3_1.pdf}}
\subfigure[Mid Term]{\includegraphics[width=0.4\linewidth]{Figure_A3_2.pdf}}
\subfigure[Far Term]{\includegraphics[width=0.4\linewidth]{Figure_A3_3.pdf}}
\caption{Figure shows similar result as shown in figure 3 using near-term, mid-term and far-term of Sardar Sarovar}
\label{FigA2}

\end{figure}

\begin{figure}
\renewcommand{\thefigure}{A4}
\centering  
\subfigure[Kadana]{\includegraphics[width=0.45\linewidth]{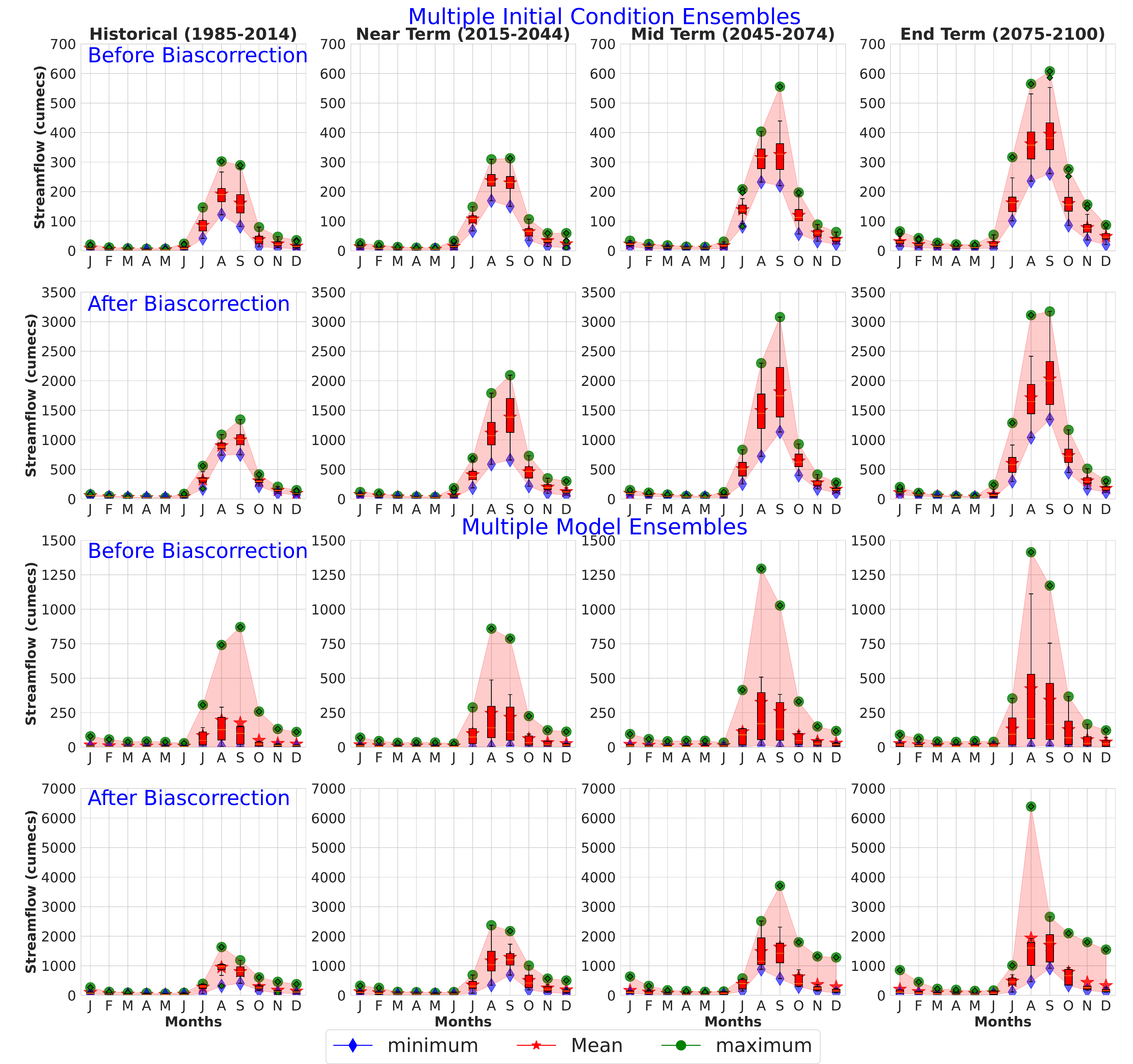}}
\subfigure[Ukai]{\includegraphics[width=0.45\linewidth]{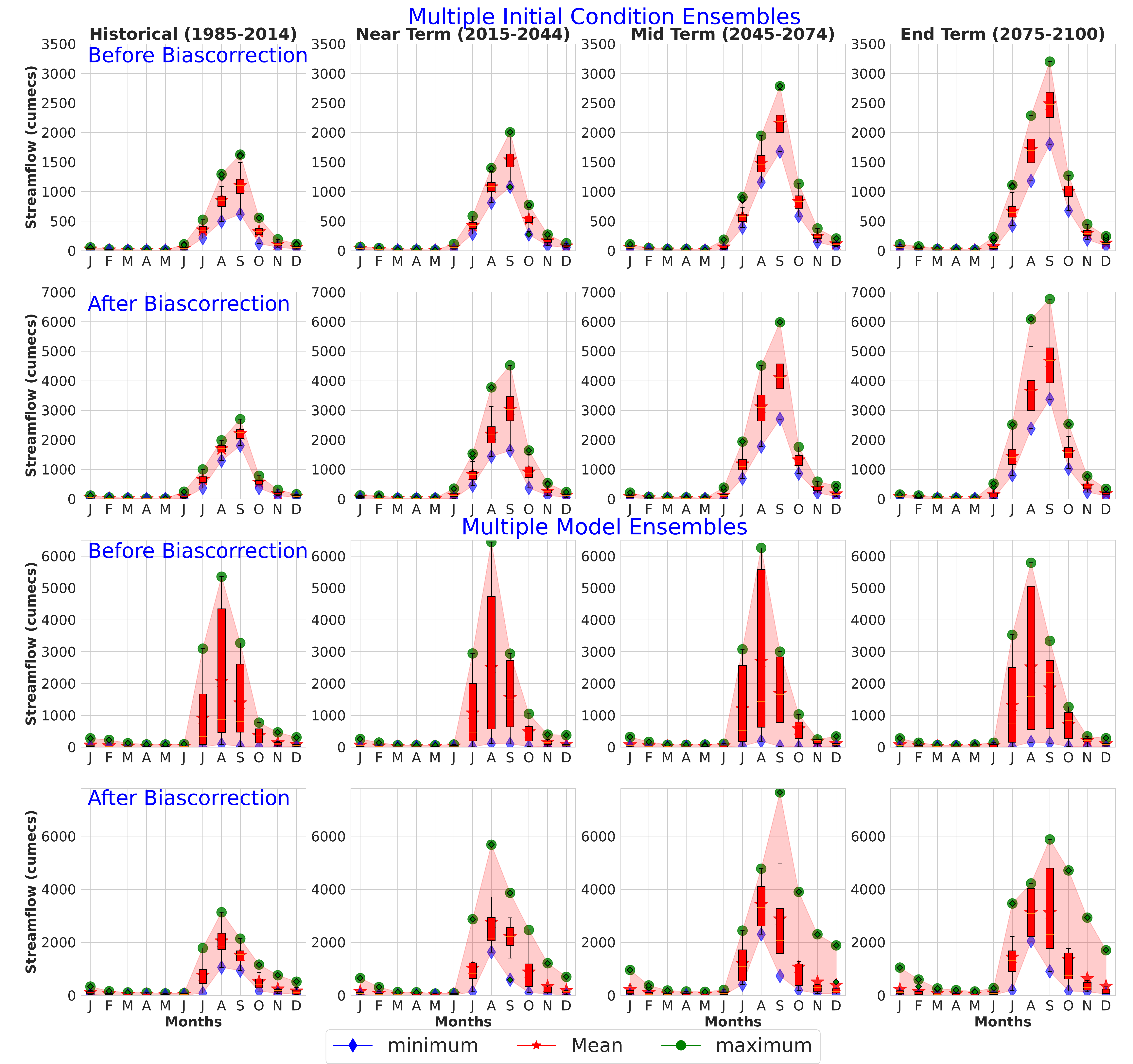}}
\caption{Figure shows similar result as shown in figure 4 for Kadana and Ukai}
\label{FigA3}
\end{figure}

\begin{figure}
\renewcommand{\thefigure}{A5}
\centering  
\subfigure[Kadana]{\includegraphics[width=0.45\linewidth]{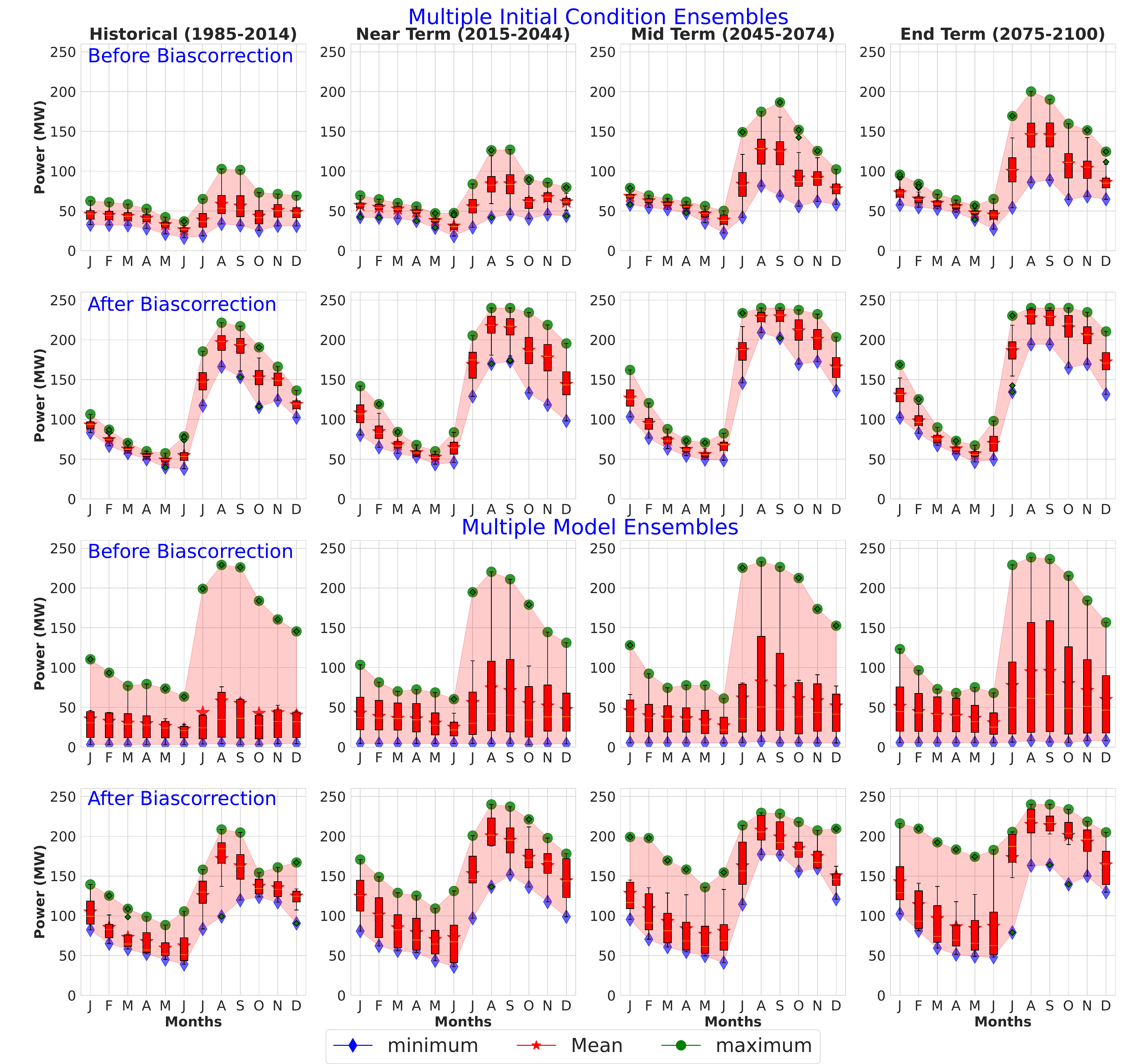}}
\subfigure[Ukai]{\includegraphics[width=0.45\linewidth]{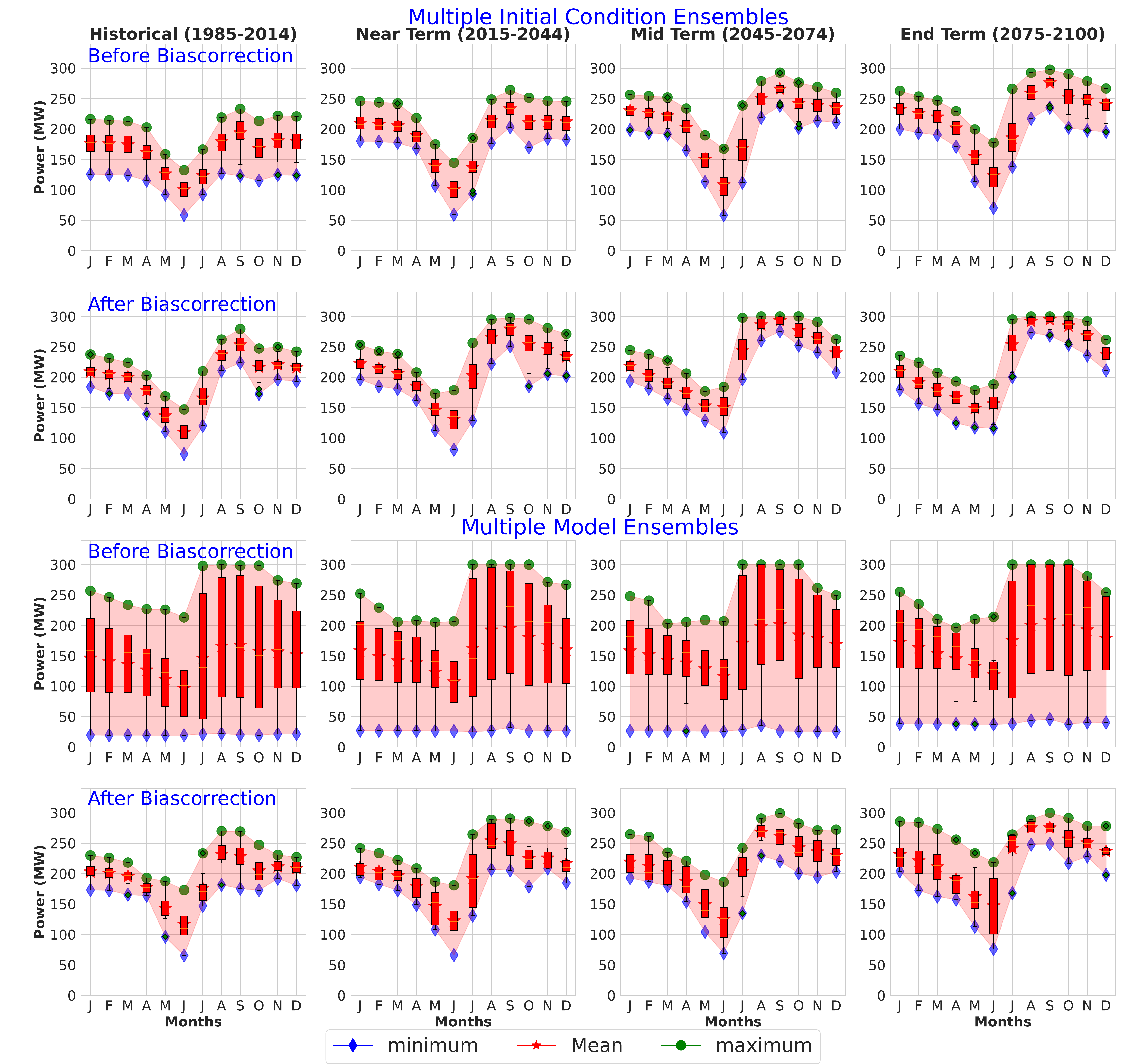}}
\caption{Figure shows similar result as shown in figure 5 for Kadana and Ukai}
\label{FigA4}
\end{figure}

\begin{figure}
\renewcommand{\thefigure}{A6}
\centering  
\subfigure[Kadana]{\includegraphics[width=0.7\linewidth]{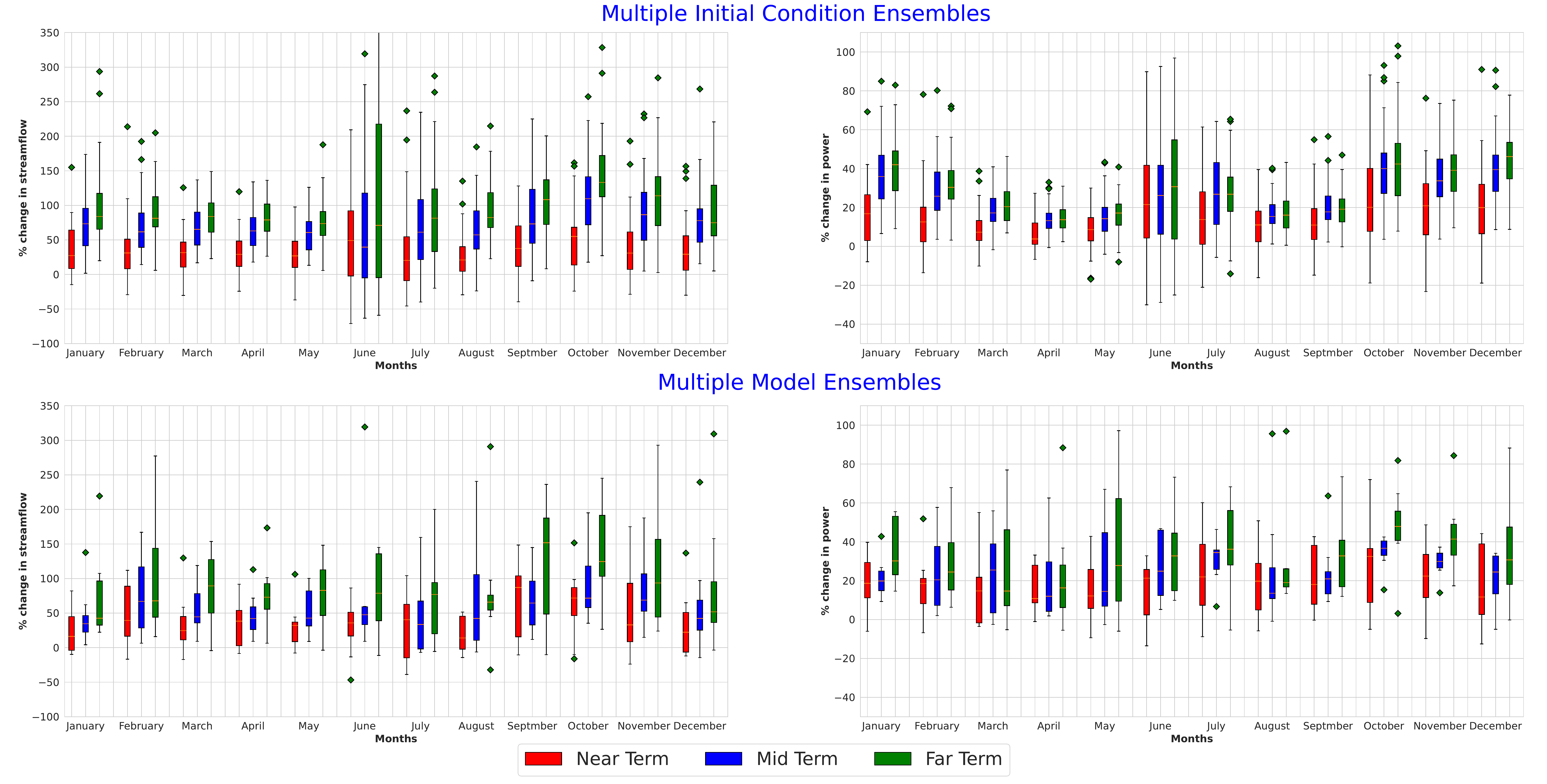}}
\subfigure[Ukai]{\includegraphics[width=0.7\linewidth]{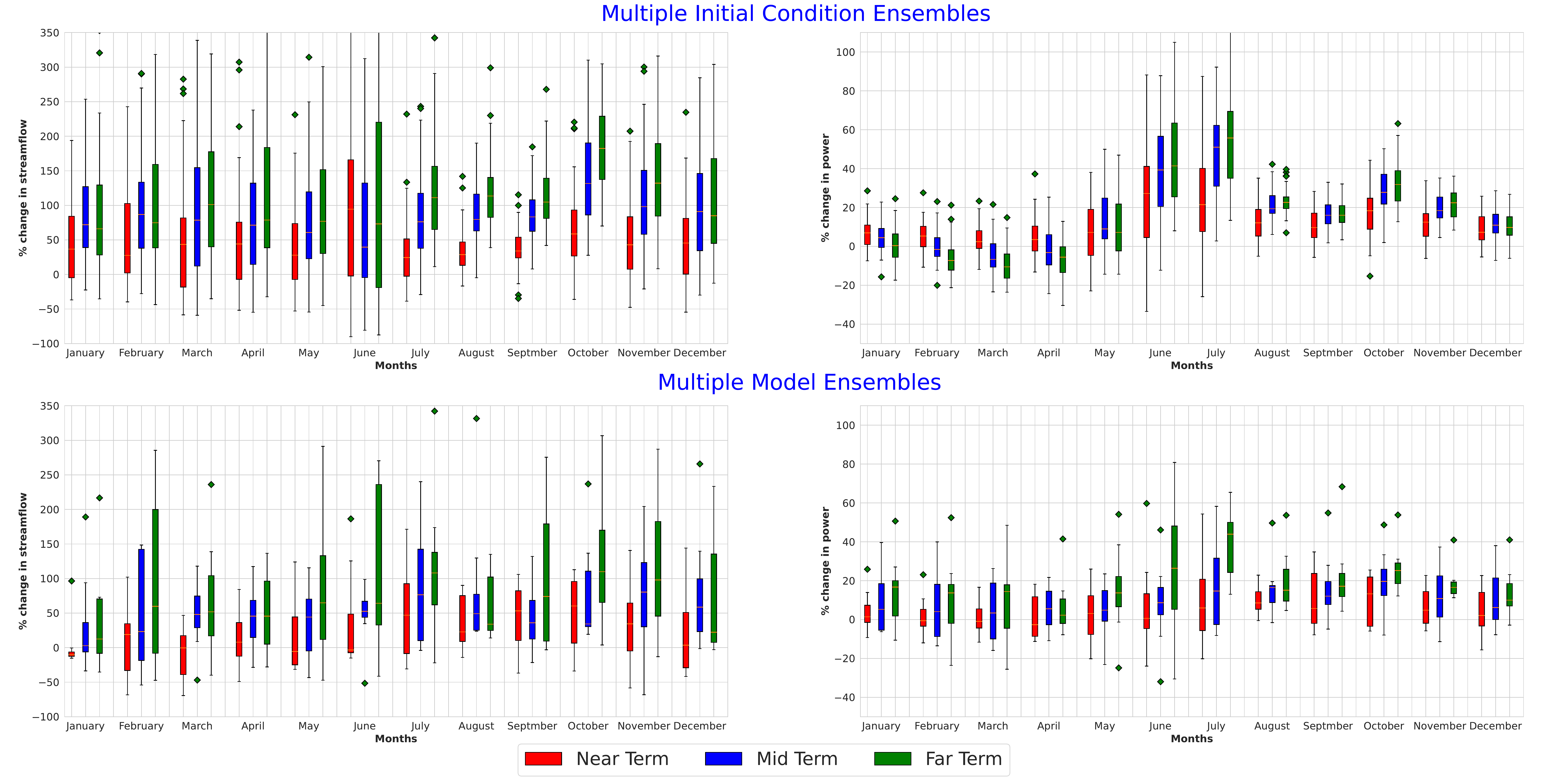}}
\caption{\textbf{Percentage change of streamflow and hydropower}: percentage change in streamflow is not reflected in percentage change in hydropower. Figure shows similar result as shown in figure 6 for Kadana and Ukai}
\label{FigA5}
\end{figure}
\newpage

\end{document}